\documentclass[english,aps,prl,twocolumn,superscriptaddress,groupedaddress,footinbib,reprint]{revtex4}
\usepackage{graphicx}
\usepackage{amsmath}
\usepackage{amssymb}
\usepackage{comment}
\usepackage{subfigure}
\usepackage{color}
\usepackage{soul}

\newcommand{\wmax}{w_{\mathrm{max}}}
\newcommand{\qin}{q_{\mathrm{in}}}

\newcommand{\qtotal}{q_{\mathrm{total}}}
\newcommand{\Qtotal}{Q_{\mathrm{total}}}
\newcommand{\qfold}{q_{\mathrm{snap}}}
\newcommand{\Qfold}{Q_{\mathrm{snap}}}
\newcommand{\dflow}{\delta_{\mathrm{flow}}}

\newcommand{\id}{\mathrm{d}} 
\renewcommand{\d}[2]{\frac{\id #1}{\id #2}} 
\newcommand{\dd}[2]{\frac{\id^2 #1}{\id #2^2}} 
\newcommand{\df}[2]{\frac{\id^4 #1}{\id #2^4}} 

\newcommand \beq{\begin{equation}}
\newcommand \eeq{\end{equation}}
\newcommand \beqn{\begin{equation*}}
\newcommand \eeqn{\end{equation*}}

\begin{document}

\title{Passive control of viscous flow via elastic snap-through}

\author{Michael Gomez}
\affiliation{Mathematical Institute, Andrew Wiles Building, University of Oxford, Woodstock Road, Oxford OX2 6GG, UK}

\author{Derek E.~Moulton}
\affiliation{Mathematical Institute, Andrew Wiles Building, University of Oxford, Woodstock Road, Oxford OX2 6GG, UK}

\author{Dominic Vella}
\email{dominic.vella@maths.ox.ac.uk}
\affiliation{Mathematical Institute, Andrew Wiles Building, University of Oxford, Woodstock Road, Oxford OX2 6GG, UK}

\begin{abstract}
We demonstrate the passive control of viscous flow in a channel by using an elastic arch embedded in the flow. Depending on the fluid flux, the arch may `snap' between two states --- constricting and unconstricting ---  that differ in hydraulic conductivity by up to an order of magnitude. We use a combination of experiments at a macroscopic scale and theory to study the constricting and unconstricting states, and determine the critical flux required to transition between them. We show that such a device may be precisely tuned for use in a range of applications, and in particular has potential  as a passive microfluidic fuse to prevent excessive fluxes in rigid-walled channels. \end{abstract} 

\makeatother
\maketitle

Elastic elements are finding increasing utility in engineering design, from aeronautics to architecture \cite{Khoo:2011wp}. The potential for passive control offered by morphable components holds particular promise in microfluidics where a library of design considerations to control the flow of fluid exists, including the geometrical, chemical and mechanical characteristics of the channel \cite{stone2004}. Of these, many are fixed at the design stage (e.g.~the network connectivity) and are difficult to change subsequently, while others can be changed actively during operation. For example, the Quake valve \citep{unger2000,oh2006} allows flow in a primary channel  to be blocked off by inflating control channels. Channel flexibility has been exploited to control flows by bending the device \citep{holmes2013}, applying a varying potential difference to create a  microfluidic pump \citep{tavakol2014} or simply by turning mechanical screws to constrict flow \citep{weibel2005}. 

The above examples have two features in common:  they are actively controlled and  generate a smoothly transitioning fluid flow. However, this active control may mean that miniaturization becomes difficult if, for example, additional power sources are required. Passive control, the ability of a flow to self-regulate, is then desirable, and has led to the development of passive pumps in microfluidic devices \cite{leslie2009,Hosoi2004}. In other circumstances, a rapid and switch-like response may also be useful, for example as a logic element in microfluidic circuits \cite{Weaver2010}, in fluidic gating \cite{Mosadegh:2010gf}, or as a fuse to limit the fluid flux within a channel to some predetermined maximum.

\begin{figure}
\centering
\includegraphics[width = \columnwidth]{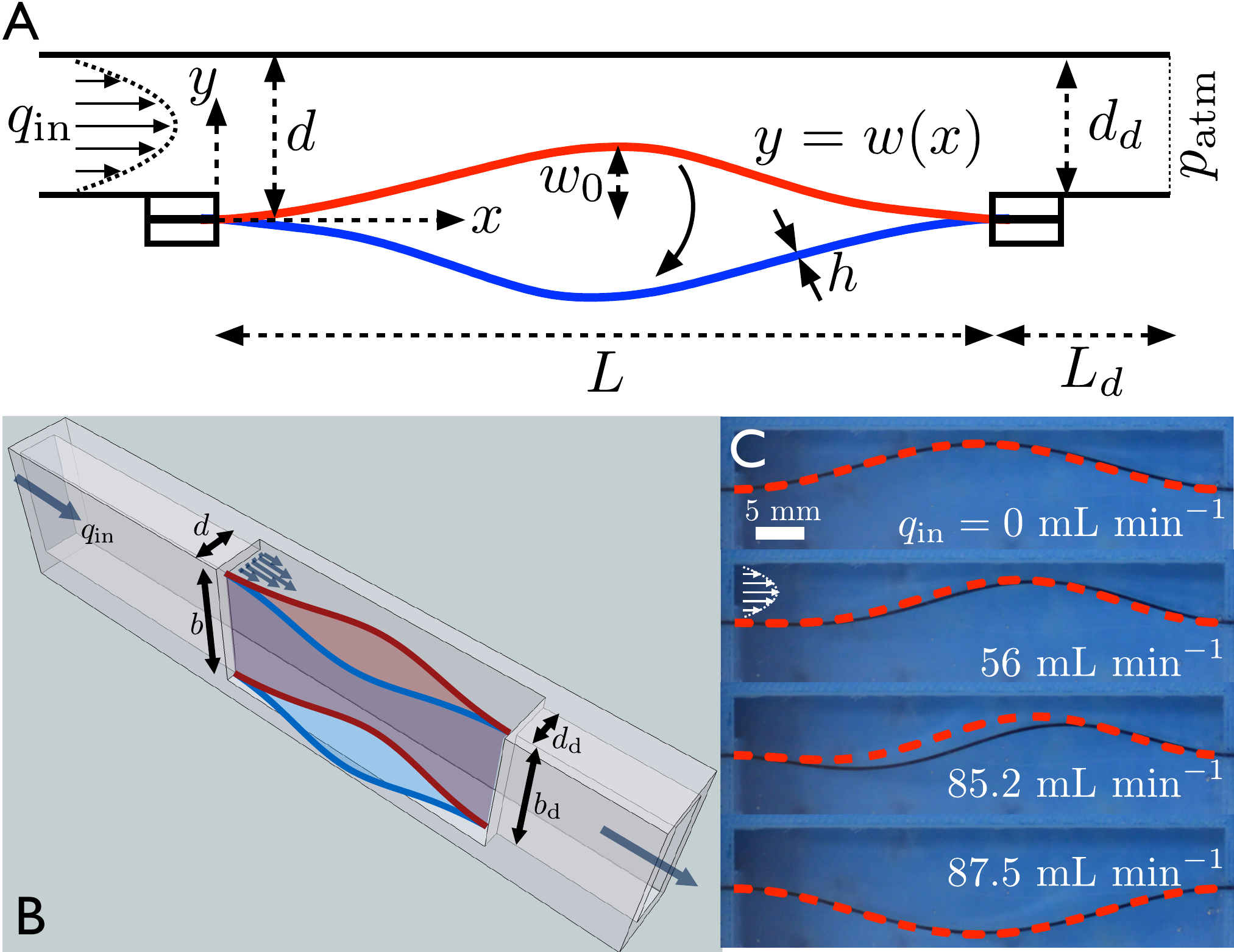}
\caption{Viscous flow through a channel containing a flexible wall. (A) A thin elastic strip, buckled into an arch, initially constricts part of a channel (red shape). At higher flow rates, the arch rapidly snaps through (blue shape); the flow is then unconstricted and  the channel's conductivity increases. (B) Three-dimensional view showing the finite channel depth. (C) Shapes of the arch during a snapping experiment ($h = 0.25~\mathrm{mm}$, $w_0 = 4.7~\mathrm{mm}$, $\eta = 1.60 \pm 0.10~\mathrm{Pa}~\mathrm{s}$), together with the shapes predicted by our beam-lubrication model (red dashed curves).}
\label{fig:setup}
\end{figure}

Elastic `snap-through', in which a system rapidly transitions from one state to another (just as an umbrella rapidly inverts in high winds) is a natural candidate for such a passive control mechanism: snap-through is generally fast, repeatable, and provides a large shape change. Snap-through has been harnessed in biology and engineering to generate fast motions between two states \citep{forterre2005,Han2004,brinkmeyer2012,holmes2007,Overvelde2015}. Previous studies have focussed on snapping due to dry, mechanical loads including indentation \citep{Pandey2014}, end rotation \citep{Gomez2017} and electrostatic forces \citep{Krylov2008}, or capillary forces in wet systems \cite{Fargette2014}. However, snap-through caused by bulk fluid flow remains relatively unexplored. Similarly, the use of elastic deformation to control fluid flows has largely focussed on the development of fluidic diodes and valves \cite{oh2006,leslie2009}.

To illustrate the mechanics of snap-through and its possible use to control  flow, we performed macroscopic experiments. Flow occurs in a channel of rectangular cross-section (width $d = 6~\mathrm{mm}$, depth $b = 23~\mathrm{mm}$) in which one of the bounding walls is replaced by a flexible strip of bi-axially oriented polyethylene terephthalate (PET) film (Young's modulus $E =  5.72 \pm 0.52~\mathrm{GPa}$).
 The rigid portion of the channel was 3D printed, with one of the walls fabricated from transparent acrylic to visualize the flow-induced deformation of the flexible element. The ends of the strip were clamped parallel to the flow direction, a distance  $L = 50~\mathrm{mm}$ apart, using  thin notches  built into the surrounding channel walls  (see fig.~\ref{fig:setup}a). The bending stiffness of the strip was varied by using different thicknesses $h \in \lbrace0.1,0.25\rbrace~\mathrm{mm}$ \cite{endnote35}.

A controlled volumetric flux, $\qin$, of glycerol (viscosity range $1.10\mathrm{~Pa\,s}\leq\eta\leq1.80~\mathrm{Pa\,s}$) was introduced using a syringe pump  (Harvard Apparatus PHD Ultra Standard Infuse/Withdraw 70-3006). Next to the arch the (reduced) Reynolds number is 
 $\mathrm{Re} = O(10^{-2})$ so that fluid inertia is negligible.
We measured the fluid pressure at the upstream end of the arch  using a voltage-output pressure transducer (OMEGA PX40-50BHG5V). We were able to accurately measure pressures larger than $140~\mathrm{Pa}$ with typical uncertainty $\pm 20~\mathrm{Pa}$ (due to uncertainties in the voltage measurement).

A key geometric parameter is the relative height of the arch in the absence of flow, $w_0$, to the upstream channel width $d$ (fig.~\ref{fig:setup}a). This arch height was varied within the channel assembly by changing the length of the strip prior to clamping. The difference between the natural length of the strip, $L_{\mathrm{strip}}$, and the horizontal distance between the two clamping points is referred to as the end--shortening $\Delta L=L_{\mathrm{strip}}-L\ll L$; for shallow arches $\Delta L$ is related to the arch amplitude by $w_0\approx2(L\,\Delta L)^{1/2}/\pi$ (using the Euler-buckling mode $w(x)=w_0\bigl[1-\cos(2\pi x/L)\bigr]/2$ \citep{holmes2013}).

At the start of each experiment, the arch was placed in a constricting state with its midpoint directed into the channel (fig.~\ref{fig:setup}a). 
To determine the dependence of the system on the fluid flux, $\qin$, this flux was ramped from zero at a rate $\dot{q}_{\mathrm{in}}=2~\mathrm{mL}~\mathrm{min}^{-2}$ (when $h = 0.1~\mathrm{mm}$) or
$\dot{q}_{\mathrm{in}}=70~\mathrm{mL}~\mathrm{min}^{-2}$ (when $h = 0.25~\mathrm{mm}$). In both cases the ratio of the convective timescale ($ L b d/\qin$) to the ramping timescale ($ \qin/\dot{q}_{\mathrm{in}}$) is $O(0.1)$ at the point of snap-through --- ramping occurs  approximately quasi-statically. A digital camera  mounted above the acrylic wall recorded the shape of the arch and allowed the midpoint height $w_0$ to be measured to an accuracy $\pm 0.2~\mathrm{mm}$. 

Snapshots of the arch shape as $\qin$ changes are shown in fig.~\ref{fig:setup}c (for movies see \cite{endnote35}). As $\qin$ increases, the shape of the arch changes only slightly at first, developing a small asymmetry due to the pressure gradient that drives the flow. However, at a critical value of $\qin$ the shape changes dramatically: the arch suddenly adopts the opposite curvature (last panel in fig.~\ref{fig:setup}c) and, if the flux $\qin$ is subsequently reduced, the arch remains in this  `snapped', unconstricting configuration. 

\begin{figure*}
\centering
\includegraphics[width = 2\columnwidth]{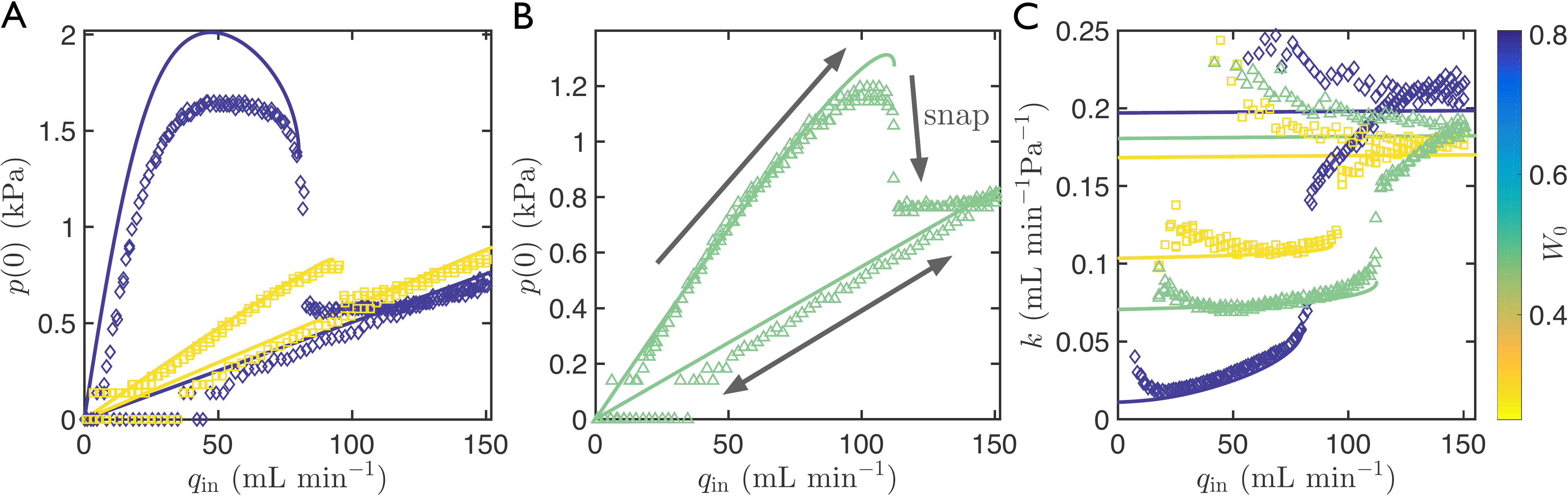}
\caption{Pressure-flux relations for a flexible channel ($h = 0.25~\mathrm{mm}$, $\eta = 1.60 \pm 0.10~\mathrm{Pa}~\mathrm{s}$). (A) Evolution of the upstream pressure, $p(0)$, for different channel blocking parameters $W_0 = w_0/d$ (indicated by the colorbar). For each $W_0$, three data sets through the snapping transition are shown, together with a fourth in which the arch remains in the snapped configuration throughout  (symbols). Predictions from the beam-lubrication model, \eqref{eqn:pressure}, are also shown (solid curves). (The snapping transition appears continuous in  experiments because the arch motion is overdamped.) (B) The hysteresis loop highlighted for intermediate $W_0$. (C) The effective hydraulic conductivity $k = \qin/p(0)$ is plotted for the same data (with $p(0)>140~\mathrm{Pa}$, to avoid noise due to inaccurate readings at low pressure.)}
\label{fig:pressures}
\end{figure*}

To quantify the behavior of this flexible channel, we measured the pressure at the upstream end of the arch, $p(0)$, as a function of the imposed flux; results for different initial arch heights are shown in fig.~\ref{fig:pressures}a. For small arch heights, the pressure increases approximately linearly with $\qin$ before snap-through, as would be expected for Poiseuille flow in a rigid channel. However, for larger arch heights, $w_0/d\nearrow1$, the contrast with Poiseuille flow becomes apparent: the pressure changes nonlinearly with $\qin$ and is even non-monotonic,  reaching a maximum prior to snapping (fig.~\ref{fig:pressures}a). Over a large range of fluxes prior to snapping, the channel therefore has a softening property whereby the effective hydraulic conductivity, which we define as $k =  \qin/p(0)$, increases smoothly with increasing flux (fig.~\ref{fig:pressures}c). 

Snap-through causes even more significant changes: the pressure drops discontinuously, even though the  flux has increased, because the channel  switches from a constricted state to an unconstricted state. The contrast between  the channel conductivities in the two states is large and grows as the arch height, $w_0$, grows (fig.~\ref{fig:pressures}c). The system exhibits hysteresis since the snapped configuration remains stable if $\qin$ is decreased (fig.~\ref{fig:pressures}b). 

\begin{figure}
\centering
\includegraphics[width = \columnwidth]{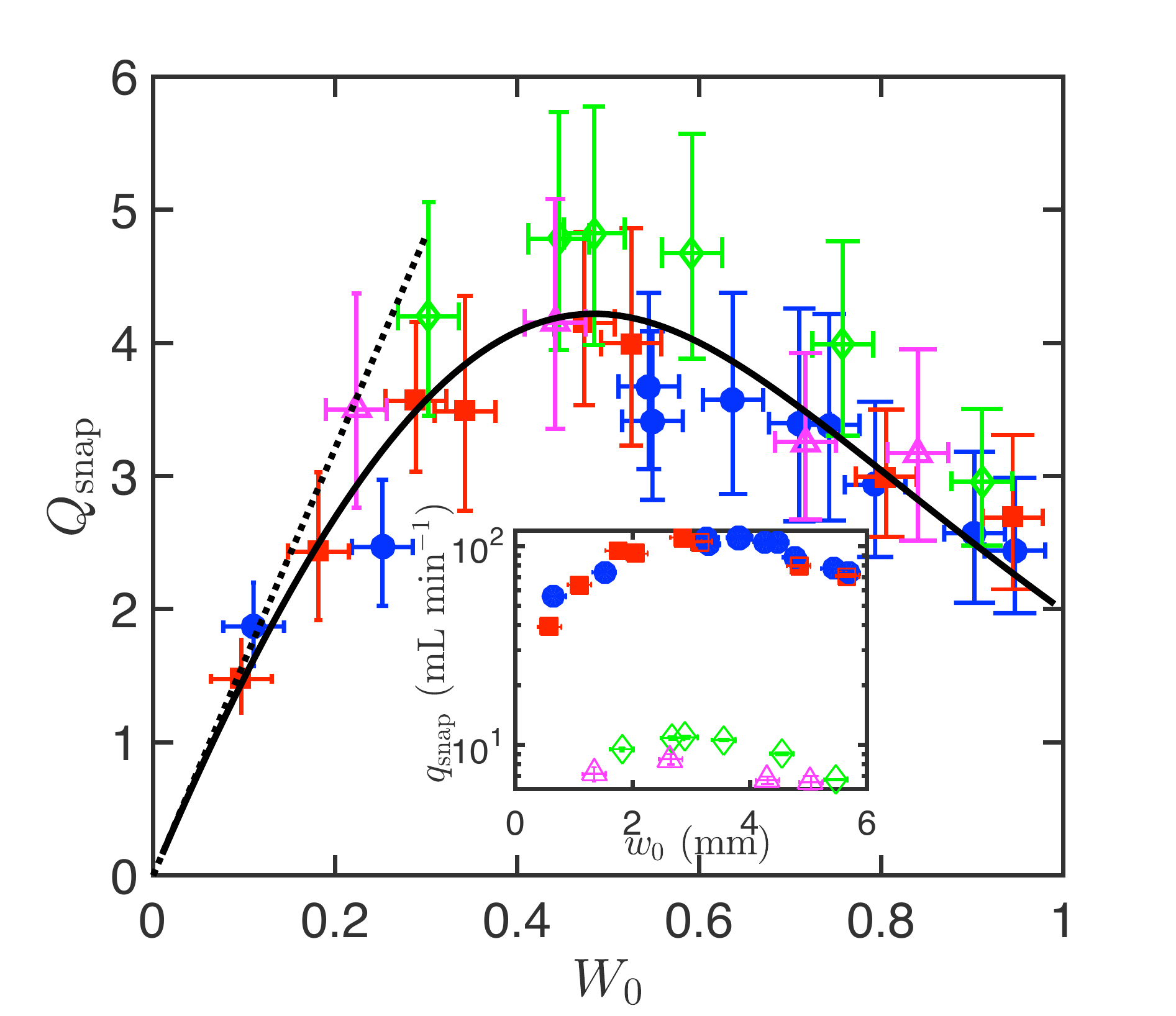}
\caption{Critical flux for snap-through. Inset: Experimentally measured snap-through flux, $\qfold$ (averaged over three runs), as a function of the initial midpoint height, $w_0$. Data is shown for $h = 0.25~\mathrm{mm}$ with $\eta = 1.38\pm 0.17~\mathrm{Pa}~\mathrm{s}$ (blue circles) and $\eta = 1.61\pm 0.18~\mathrm{Pa}~\mathrm{s}$ (red squares); and for $h = 0.1~\mathrm{mm}$ with  $\eta = 1.20\pm 0.10~\mathrm{Pa}~\mathrm{s}$ (green diamonds) and $\eta = 1.33\pm 0.08~\mathrm{Pa}~\mathrm{s}$ (magenta triangles;  increasing $q_{\mathrm{in}}$  in steps of $0.25~\mathrm{mL}~\mathrm{min}^{-1}$ every minute rather than ramping). Horizontal error bars correspond to the $\pm 0.2~\mathrm{mm}$ uncertainty  in $w_0$; vertical error bars give the standard deviation of the measured values. Main plot: Rescaling to plot the dimensionless flux $\Qfold=\eta L^5\qfold/(Bbd^4)$ in terms of the channel blocking parameter $W_0=w_0/d$, the data collapse onto the prediction of our numerical analysis (solid black curve). Vertical error bars here also account for uncertainties in the bending stiffness $B$ and viscosity $\eta$. Also plotted is the asymptotic result $\Qfold\approx 16W_0$ valid for $W_0 \ll 1$  \cite{endnote35} (black dotted line).}
\label{fig:snappingflux}
\end{figure}

A key quantity of interest is the critical flux, $\qfold$, at which snap-through occurs; fig.~\ref{fig:snappingflux} (inset) shows that this depends not only on the arch height, $w_0$, but also on the flexibility of the arch and the liquid's  properties. Surprisingly, we find that the value of $\qfold$ is a non-monotonic function of arch height: for given material parameters,  a maximum value of $\qfold$ is obtained at $w_0/d \approx 0.5$.

To gain theoretical insight we first note that the deflection $\delta$ of an elastic strip, of length $L$ and bending stiffness $B$, due to a  force $F$ (per unit length) scales as $\delta\sim FL^3/B$ \cite{Timoshenko}.
Here the typical force  $F\sim p L$, where $p$ is the fluid pressure, and hence the induced deformation $\dflow\sim pL^4/B$. The Poiseuille law \cite{Happel} for the pressure drop along a slender channel of width $d$ and depth $b$, with an obstruction of maximum size $\wmax$, suggests that $p \sim \eta L\qin/[b(d-\wmax)^3]$. This pressure estimate then gives $\dflow\sim \eta L^5\qin/[Bb(d-\wmax)^3]$, which may be compared with the initial arch height $w_0$ to estimate the threshold flux for snap-through  (analogously to point indentation \cite{Pandey2014}) as 
\begin{equation}
\qfold\sim\frac{Bb\left(d-\wmax\right)^3}{\eta L^5}w_0.
\end{equation}
This may be written in terms of the channel blocking parameter, $W_0=w_0/d$, as 
\begin{equation}
\Qfold\sim W_0 \left(1-\frac{\wmax}{d}\right)^3,
\label{eqn:QfoldScale}
\end{equation}
where a dimensionless fluid flux is 
\begin{equation}
Q = \frac{\eta L^{5} }{B b d^4}\qin. \label{eqn:defnQ}
\end{equation} 
This non-dimensionalization provides an excellent collapse of the experimental data onto a single master curve (fig.~\ref{fig:snappingflux}). Moreover, the non-monotonic behavior observed in fig.~\ref{fig:snappingflux} is qualitatively explained by  \eqref{eqn:QfoldScale}: for small channel blocking parameter, $W_0=w_0/d\ll1$, the maximum arch displacement $\wmax \ll d$, and hence $\Qfold \sim W_0$. However, when $\wmax$ becomes comparable to the channel width $d$ ($W_0\nearrow1$), $\Qfold$ decreases. 

To go beyond these scaling arguments, we  formulate a model coupling the shape of the arch with the fluid pressure by exploiting the thin-film geometry and the shallow slope of the arch.  This allows us to use the one-dimensional linear beam equation \cite{Howell2009}
\begin{equation}
 B \df{w}{x} +  T\dd{w}{x} + p(x) = 0, \qquad 0 < x < L, \label{eqn:beameqn}
\end{equation} 
to describe the transverse displacement, $w(x)$, of the arch, with $T$ the compressive force in the arch, and $p(x)$ the hydrodynamic pressure. (An analysis of the shear stress exerted on the arch by the fluid shows \cite{Kodio2017} that the compressive force $T$ is spatially uniform provided that $|\id w/\id x|\ll1$, as already assumed in using the linear beam equation.) Assuming that the strip is inextensible \cite{Pandey2014}, the imposed end-shortening $\Delta L$ leads to the  constraint   
\begin{equation}
\int_0^L\left(\d{w}{x}\right)^2~\id x=2\Delta L. \label{eqn:end-shortening}
\end{equation} 
The ends of the arch, at $x = 0$ and $x = L$, are clamped i.e.~$w(0) = w'(0) = w(L) = w'(L) = 0$ (with primes denoting differentiation with respect to $x$).

To determine the pressure within the liquid, $p(x)$, we use lubrication theory \cite{Leal2007}, consistent with our assumption of small slopes, $|\id w/\id x| \ll 1$. Using standard methods, the pressure may be expressed \cite{endnote35} as
\begin{equation}
p(x) = p(L) + \frac{12\eta q_{\mathrm{in}}}{b}  \int_x^{L} \frac{K(w(\xi))}{\left[d-w(\xi)\right]^{3}}~\mathrm{d}\xi, \label{eqn:pressure}
\end{equation} 
where we use a geometric correction factor \citep{stone2004,Happel}
\begin{equation*}
K(w) = \left[1-6 \left(\frac{2}{\pi}\right)^5 \frac{d - w}{b}\right]^{-1},
\end{equation*}
to account for the finite depth of the channel. The pressure at the downstream end of the arch depends on the downstream geometry of the channel (denoted with subscript $d$, as in fig.~\ref{fig:setup}a) and is given by
\begin{equation*}
p(L) = \frac{12\eta q_{\mathrm{in}}L_{\mathrm{d}}}{b_{\mathrm{d}} d_{\mathrm{d}}^3} \left[1-6 \left(\frac{2}{\pi}\right)^5 \frac{d_{\mathrm{d}}}{b_{\mathrm{d}}}\right]^{-1}, \label{eqn:downstreampressure}
\end{equation*}
measured relative to the ambient pressure (which is imposed at the end of the channel, $x = L + L_{\mathrm{d}}$). 

We introduce the dimensionless variables $X = x/L$, $W = w/d$, and $P = p/p^{*}$ where $p^*=Bd/L^4$ is the pressure scale introduced by the beam equation \eqref{eqn:beameqn}. With this non-dimensionalization, there are two key governing parameters: the dimensionless flux $Q$, defined in \eqref{eqn:defnQ}, and the  channel blocking parameter $W_0= w_0/d \approx 2(L\,\Delta L)^{1/2}/(\pi d)$. 

The dimensionless versions of equations \eqref{eqn:beameqn}--\eqref{eqn:pressure} may be solved for given values of $W_0$ and $Q$ to determine both the arch shape and the dimensionless pressure field, $P(X)$. Predicted arch shapes are shown in fig.~\ref{fig:setup}c, superimposed on the experimentally observed shapes;  the agreement between theory and experiment is very good for all values of $Q$ investigated, including  beyond the snap-through transition. The discrepancy is largest close to snap-through (third panel of fig.~\ref{fig:setup}c), since the sensitivity to the precise value of $Q$ is largest here. The predicted (dimensional) upstream pressure $p(0)$ is shown in fig.~\ref{fig:pressures}a,b, with corresponding conductivities $k = \qin/p(0)$ plotted in fig.~\ref{fig:pressures}c; both generally agree well with experiment (errors in the conductivity at low fluxes are  due to uncertainties in the measurement of low pressures). Close to total blocking, $W_0\approx1$, there is  a systematic error in the model, which we attribute to the relatively large arch slopes at the midpoint that are not captured by our use of lubrication and linear beam theories.  Nevertheless, the model captures the qualitative behavior of the pressure throughout, including the non-monotonicity of $p(0)$ as a function of $\qin$.

A numerical analysis of the problem shows \cite{endnote35} that the snap-through transition is a saddle-node bifurcation: the constricting state ceases to exist at a critical value $Q=\Qfold$ without first becoming unstable \cite{Pandey2014}. The numerically determined value of $\Qfold(W_0)$ reproduces the experimentally determined master curve; see fig.~\ref{fig:snappingflux}. For $W_0\ll1$, an asymptotic analysis shows that $\Qfold\approx 16W_0$, reproducing the linear scaling of \eqref{eqn:QfoldScale}. For $0.1\lesssim W_0\leq1$, we find that $\Qfold$ varies by less than a factor of 2, with $2\lesssim\Qfold\lesssim4$.

The system we have presented is irreversible --- post snapping the strip cannot return to the constricting state without direct intervention. However, this is not a fundamental feature: reversibility may be accomplished by introducing flow in an access channel to the region below the arch, to snap the arch back to its original position (see e.g.~\cite{Mosadegh:2010gf}).  Alternatively, an automatic reset, which may be desirable in some applications, may be easily achieved by clamping one end of the arch at an angle to the horizontal \cite{Gomez2017,Arena2017} so that the snapped configuration is not in equilibrium in the absence of flow. In this case, the system exhibits a hysteresis loop with an increase in the input flux generating a snap in one direction, and a subsequent (further) decrease in flux causing a snap back (see fig.~S6 of \cite{endnote35}).

\begin{figure}
\centering
\includegraphics[width = \columnwidth]{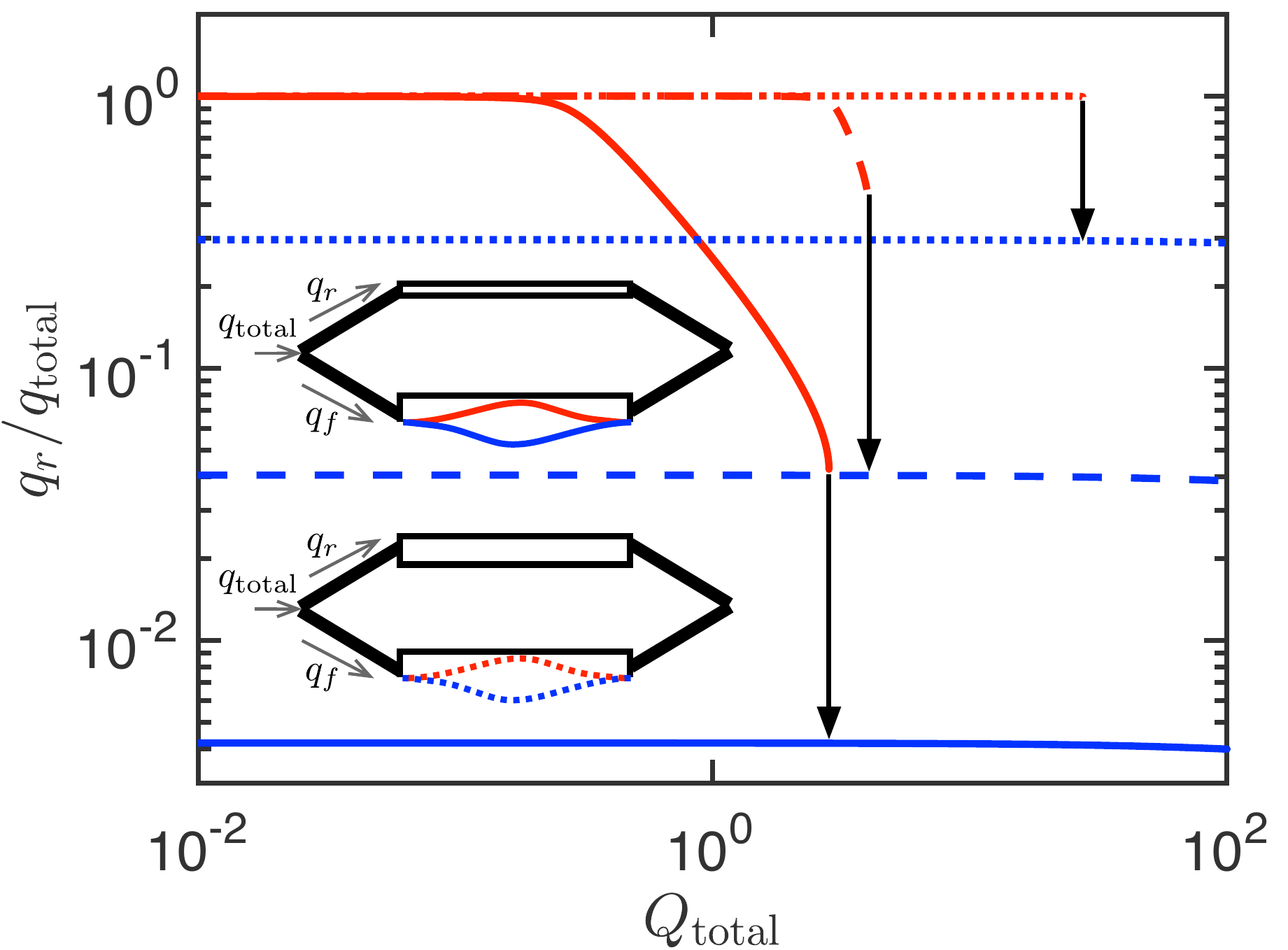}
\caption{Flow limiting using snap-through. Inset: A channel with a flexible wall is connected in parallel to a rigid channel of constant conductivity (schematics not drawn to scale). Main figure: For low total fluxes $\Qtotal =  \eta L^5\qtotal/(Bb d^4)$ with the flexible channel constricting (red curves), almost all of the flow is directed through the rigid channel, i.e.~$q_r/\qtotal \approx 1$. This is diverted through the flexible channel as soon as the arch snaps (blue curves). Here $W_0=0.99$ and numerical results are shown for different conductivity ratios $\lambda$ between the channels: $\lambda = 10^{-2}$ (solid curves), $\lambda = 10^{-1}$ (dashed curves) and $\lambda = 1$ (dotted curves). }
\label{fig:Fuse}
\end{figure}

In both the irreversible and reversible scenarios, the quantitative features of the mechanism (e.g.~the critical snapping fluxes and the corresponding  change in conductivity) may be precisely tuned. Therefore, with an arch element coupled to other components, a range of design possibilities opens up.  For example, in fig.~\ref{fig:Fuse} we demonstrate the potential for a passive fluid `fuse'.  Here we have placed an arch element in parallel with  another, entirely rigid, channel (fig.~\ref{fig:Fuse} inset).  Denoting the (constant) effective hydraulic conductivity of the rigid channel by $k_r$, and the (variable) conductivity of the flexible channel by $k_f(q_f)$, the ratio of the fluxes through each of the two channels is $q_r/q_f=k_r/k_f$ by the Poiseuille law.

Denoting the total flux $\qtotal=q_f+q_r$ and calculating $q_r/\qtotal$, the fraction of the total flux that passes through the rigid channel, we find a switch-like response (fig.~\ref{fig:Fuse}): while the arch is in a constricting shape, most of the fluid passes through the rigid channel, but once the arch snaps, much of the fluid is diverted to the now unconstricted flexible channel. The rigid channel is effectively `short-circuited'. The efficiency of the fuse, defined as the decrease in $q_r$ caused by snap-through divided by its value prior to snap-through, may be tuned by varying the geometric parameters of each channel \cite{endnote35}.

We have shown at a laboratory scale that the pressure gradient associated with a viscous flow can be used to cause snap-through of an embedded elastic element. The system considered has a number of novel flow properties including a highly nonlinear pressure-flux relationship, discontinuous conductivity and history dependence. These properties may find application in microfluidic systems such as cell-sorting or, as we have shown, provide a means to protect microfluidic systems from high fluxes. Similarly, the discontinuous transition we observe is similar to that seen in capillary burst valves \cite{Cho2007} and gas release valves  \cite{Kim2009}.  A simple analysis \cite{endnote35} shows that when scaling down to the microscale, the expected range of snap-through fluxes are well within experimentally obtainable values.  For such applications our study thus provides a first analysis of flow-induced snapping and guidance for choosing material parameters to tune the critical flux. While viscous flow control is readily applicable to microfluidics, the passive control and rapid transition capabilities of elastic materials is increasingly being exploited more broadly, e.g.~in soft robotics and morphing skins \cite{Overvelde2015,Thill:2008uk}. Developing theoretical models that provide intuition and facilitate device optimization will be critical in these burgeoning fields of technology.


\begin{acknowledgments}
The research leading to these results has received funding from the European Research Council under the European Union's Horizon 2020 Programme / ERC Grant Agreement no.~637334 (DV) and the EPSRC Grant No. EP/ M50659X/1 (MG). We are grateful to John Wettlaufer for encouragement and helpful suggestions in the early phases of this work, Cl\'{e}ment le Gouellec for experiments in a related system, Alain Goriely for 3D printing, Chris MacMinn for laser cutting, and the John Fell Oxford University Press (OUP) Research Fund (award number 132/012). The experimental and numerical data used to generate the plots within this paper are available from http://dx.doi.org/10.5287/bodleian:VYN5z8rjr.
\end{acknowledgments}


\widetext
\clearpage
\begin{center}
\textbf{\large Supplementary information for ``\emph{Passive control of viscous flow via elastic snap-through}"}
\end{center}
\setcounter{secnumdepth}{2}
\setcounter{equation}{0}
\setcounter{figure}{0}
\setcounter{table}{0}
\setcounter{page}{1}
\makeatletter
\renewcommand{\theequation}{S\arabic{equation}}
\renewcommand{\thefigure}{S\arabic{figure}}
\renewcommand{\bibnumfmt}[1]{[S#1]}
\renewcommand{\citenumfont}[1]{S#1}
This supplementary information gives further details on the experimental setup and theoretical analysis referred to in the main text. In \S\ref{sec:expt} we provide specific details on the materials and procedures used in our experiments. In \S\ref{sec:theory} we derive the beam-lubrication model used to describe the shape of the arch as it deforms in response to fluid flow, and discuss its non-dimensionalization. In \S\ref{sec:equilibrium} we then present the bifurcation diagram of the equilibrium shapes, and perform a perturbation analysis for the case of small channel blocking parameter. In \S\ref{sec:reversible} we consider a slight modification to the boundary conditions on the arch that allows us to obtain a reversible snap-through. In \S\ref{sec:fuse} we discuss the case of a channel containing an arch element placed in parallel with a rigid channel, and analyze the fuse-like behavior that may be obtained. In \S\ref{sec:scale} we discuss the scalability of snap-through for microfluidic devices. Finally, \S\ref{sec:movie} gives details of Supplementary Movies $1$--$2$ showing the evolution of the arch shape during snapping experiments. 

\section{Experimental details}
\label{sec:expt}
We prepared strips of bi-axially oriented polyethylene terephthalate (PET) film (Goodfellow, Cambridge; density $\rho_s = 1.38~\mathrm{g}~\mathrm{cm}^{-3}$) with different thicknesses $h \in \lbrace0.1,0.25\rbrace~\mathrm{mm}$. The Young's modulus was measured  by examining vibrations of the dry arch \cite{Gomez2017SI} and found to be  $E =  5.72 \pm 0.52~\mathrm{GPa}$. Rather than bonding the strip to the channel to clamp its ends, we instead inserted the ends of the strip into thin notches (thickness $0.4~\mathrm{mm}$) built into the surrounding channel walls, parallel to the flow direction. This allowed us to easily replace the arch and hence vary its parameters between different experiments. Externally applied spring clamps ensured that the strip was effectively clamped. The remainder of the channel was 3D printed (Makerbot, Replicator $5$) and is effectively rigid. Downstream of the arch, the channel has a uniform rectangular cross-section of width $d_\mathrm{d} = 5~\mathrm{mm}$, depth  $b_\mathrm{d} = 21~\mathrm{mm}$ and length $L_\mathrm{d} = 22.2~\mathrm{mm}$.
 
The strips were laser cut so that their depth was slightly less than the channel depth $b = 23~\mathrm{mm}$ ($\approx 0.2~\mathrm{mm}$ gap), allowing the strip to move with minimal friction from the walls while minimizing any leakage. The combined effects of leakage and gravity were (further) minimized by immersing the channel in a bath of liquid.  The downstream end of the channel (at $x = L + L_d$) and the fluid below the strip (i.e.~outside the channel) remained at ambient pressure.

The working liquid is glycerol (supplied by Better Equipped; density $\rho_f =1.26\pm0.01~\mathrm{g}~\mathrm{cm}^{-3}$). Due to variations in the glycerol viscosity (from absorption of water and temperature changes), we measured the viscosity before and after each snapping experiment. All measured values are in the range $1.10\mathrm{~Pa\,s}\leq\eta\leq1.80~\mathrm{Pa\,s}$ with maximum uncertainty $\pm 0.20 ~\mathrm{Pa\,s}$.  Next to the arch, the (reduced) Reynolds number is $\mathrm{Re} = (d/L)^2 \rho_f u L/\eta$ where $u \sim \qin/(bd)$ is the incoming fluid velocity. Throughout our experiments $\qin \lesssim 150~\mathrm{mL}~\mathrm{min}^{-1}$, which gives $\mathrm{Re} = O(10^{-2})$. 

The syringe pump (Harvard Apparatus PHD Ultra Standard Infuse/Withdraw 70-3006) was loaded with two $150~\mathrm{mL}$ syringes, connected to the channel using flexible tubes (inner diameter $8~\mathrm{mm}$). Near the inlet upstream, the channel was designed to smoothly transition to the rectangular cross-section to ensure the flow was well-developed as it passes the arch. The voltage-output pressure transducer (OMEGA PX40-50BHG5V) was connected to the channel at the upstream end of the arch by $3~\mathrm{mm}$ air tubes. To obtain repeatable readings, we found it necessary to shorten the length of the air tubes to around $1~\mathrm{cm}$. The voltage readings generated by the pressure transducer were output to an Arduino and analyzed using a custom \textsc{matlab} script. By first calibrating the output voltages using flow in a uniform channel with known pressure, we were able to accurately measure pressures larger than $140~\mathrm{Pa}$.

At the start of each experiment, the channel was flushed with fluid (at low flux) to remove any air bubbles. The rate at which the flux was ramped ($\dot{q}_{\mathrm{in}}=2~\mathrm{mL}~\mathrm{min}^{-2}$ with $h = 0.1~\mathrm{mm}$, and $\dot{q}_{\mathrm{in}}=70~\mathrm{mL}~\mathrm{min}^{-2}$ with $h = 0.25~\mathrm{mm}$) was sufficiently slow that the system remained quasi-static but fast enough that snap-through occurred before the syringes were emptied; we have checked that changing $\dot{q}_{\mathrm{in}}$, or increasing the flux in small steps instead, does not change the results. A digital camera (Nikon D7000) mounted above the acrylic wall recorded the shape of the arch at $1$--$2$ second intervals. When the strip snapped, the precise point of snapping was defined to be the first instant at which the midpoint decreased below the line of zero displacement. 

\section{Theoretical formulations\label{sec:theory}}
\subsection{Elasticity}
\begin{figure}
\centering
\includegraphics[scale = 1.4]{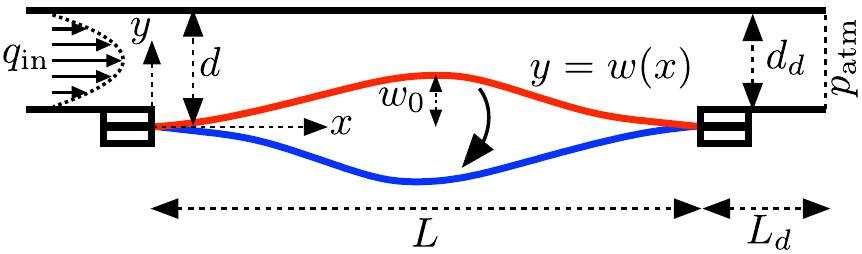} 
\caption{Cross-section of the set-up involving an elastic arch arranged next to a lubrication-type flow.}
\label{fig:schematic}
\end{figure}

A schematic of the arch in the channel is shown in figure \ref{fig:schematic}. We take Cartesian coordinates in the plane perpendicular to the depth of the arch: $x$ measures the distance downstream from the upstream end of the arch, while the $y$-direction is along the channel width. The properties of the arch are its thickness $h$, depth $b$, natural length $L_{\mathrm{strip}}$, and bending stiffness $B = Eh^3/12$ ($E$ is the Young's modulus). Note that the Poisson ratio does not appear in the expression for $B$ because we are considering a narrow strip of material (i.e. $b \ll L_{\mathrm{strip}}$) rather than an infinite plate; see for example \cite{audoly}. The midpoint height of the arch in the absence of any flow is denoted by $w_0$. The ends of the arch are clamped parallel to the flow direction a distance $L < L_{\mathrm{strip}}$ apart. Next to the arch, the channel has depth $b$ and its width is $d$ when the arch is flat ($w_0 = 0$).

To model the shape of the arch we use linear beam theory. This is justified by having a small arch thickness $h/L_{\mathrm{strip}} \ll 1$, which guarantees that the strains remain small, and ensuring the shape of the arch remains shallow; for a slender channel with $d \ll L$, this will be valid whenever we restrict $w_0 < d$.  We also assume that the depth of the arch is much larger than its thickness ($h \ll b$), so that we may neglect out-of-plane bending and twist \citep{patricio1998}. Under quasi-static loading conditions, the profile of the arch can then be written as $y = w(x)$ where $w$ is the transverse displacement (figure \ref{fig:schematic}). Performing a vertical force balance yields (see \citep{howell}, for example)
\beq
B \df{w}{x} +  T\dd{w}{x} + p = 0,  \qquad 0 < x < L,  \label{eqn:beamdim} 
\eeq
where $T$ is the unknown compressive force (per unit depth) applied to the ends of the arch, and $p$ is the fluid pressure. A horizontal force balance shows that for the shallow arch shapes considered here, spatial variations in the compressive force (due to viscous shear stresses) are negligible, so that $T$ is constant \citep{Kodio2017SI}.

We neglect the effects of extensibility, considering only arch shapes that are well past the Euler-buckling threshold under the imposed end-shortening $\Delta L = L_{\mathrm{strip}} - L$. In practice, this is satisfied simultaneously with the requirement $w_0 < d$ (needed for a shallow shape) by having a small thickness $h \ll L_{\mathrm{strip}}$; see \cite{Pandey2014SI} for a discussion of this in a related problem. The imposed end-shortening then becomes 
\beqn
\int_0^{L_{\mathrm{strip}}}\cos\theta~\id s = L,
\eeqn
where $\theta$ is the angle between the strip and the $x$-direction and $s$ is the arclength. In linear beam theory we have $s \approx x$, $L _{\mathrm{strip}}\approx L$ and $\theta \approx \id w/\id x \ll 1$ so that this constraint is approximated as
\beq
\int_0^L\left(\d{w}{x}\right)^2~\id x=2\Delta L. \label{eqn:constraintdim}
\eeq
The boundary conditions at the clamped ends are (here and throughout $'$ denotes differentiation)
\beq
w(0) = w'(0) = w(L)=w'(L) = 0. \label{eqn:clampbcsdim}
\eeq

\subsection{Lubrication theory}
We assume that the reduced Reynolds number (the relevant parameter measuring the ratio of inertia to viscosity in the slender geometry) is small. As we control the volumetric flux, $q_{\mathrm{in}}$, the typical horizontal velocity in the channel next to the arch is $u \sim q_{\mathrm{in}}/A$, where $A \sim bd $ is the cross-section area. The requirement of small reduced Reynolds number then becomes
\beqn
\delta^2 \mathrm{Re} \equiv \delta^2\frac{\rho_f (q_{\mathrm{in}}/A)L}{\eta} \ll 1,
\eeqn
where $\delta = d/L \ll 1$ is the aspect ratio of the channel, $\rho_f$ is the fluid density, and $\eta$ is the dynamic viscosity. Under this assumption, we may model the thin-film flow in the channel using lubrication theory \citep{ockendon}. Note that this is consistent with our use of linear beam theory to describe the arch shape: this assumes that the arch shape is shallow, and hence the length scale over which the channel geometry varies is much larger than its typical width. 

In the lubrication approximation, the flow in the channel is purely downstream to leading order, and the pressure gradient, $\id p/\id x$, is related to the flux $q_{\mathrm{in}}$ by the Poiseuille law. Across the length of the arch, i.e.~for each $0 < x < L$, the channel is locally rectangular with width $[d-w(x)]$ and depth $b \gg d$. This gives 
\beq
-\frac{b[d-w(x)]^3}{12 \eta} K(w(x))^{-1}\frac{\mathrm{d} p}{\mathrm{d}x} = q_{\mathrm{in}}, \quad 0 < x < L, \label{eqn:poiseuillelaw1}
\eeq
where we account for the leading-order effects of a finite depth using the correction factor \cite{stone2004SI}
\beqn
K(w) = \left[1-6 \left(\frac{2}{\pi}\right)^5 \frac{d - w}{b}\right]^{-1}.
\eeqn
Further downstream, the channel is uniform with constant width $d_{\mathrm{d}}$ and depth $b_{\mathrm{d}}$. Here we instead have 
\beq
-\frac{b_{\mathrm{d}} d_{\mathrm{d}}^3}{12 \eta} \left[1-6 \left(\frac{2}{\pi}\right)^5 \frac{d_{\mathrm{d}}}{b_{\mathrm{d}}}\right]\frac{\mathrm{d} p}{\mathrm{d}x} = q_{\mathrm{in}}, \quad x > L. \label{eqn:poiseuillelaw2}
\eeq
The end of the channel at $x = L +  L_{\mathrm{d}}$ is assumed to remain at ambient pressure. We integrate \eqref{eqn:poiseuillelaw1}--\eqref{eqn:poiseuillelaw2} to obtain the fluid pressure
\beq
p(x) = p(L) + \frac{12\eta q_{\mathrm{in}}}{b}  \int_x^{L} \frac{K(w(\xi))}{\left[d-w(\xi)\right]^3}~\mathrm{d}\xi,  \quad 0 < x < L, \label{eqn:pressureSI}
\eeq
where the pressure at the downstream end of the arch (relative to ambient) is
\beq
p(L) = \frac{12\eta q_{\mathrm{in}}L_{\mathrm{d}}}{b_{\mathrm{d}} d_{\mathrm{d}}^3} \left[1-6 \left(\frac{2}{\pi}\right)^5 \frac{d_{\mathrm{d}}}{b_{\mathrm{d}}}\right]^{-1}. \label{eqn:downstreampressureSI}
\eeq

\subsection{Non-dimensionalization}
To render the problem dimensionless, we scale the horizontal coordinate by the length $L$ between the clamped ends, i.e~we set $X=x/L$.  We choose the vertical length scale to be the upstream channel width $d$, so we introduce the dimensionless displacement $W = w/d$. The beam equation \eqref{eqn:beamdim} provides the natural pressure scale $p^{*} = B d/L^4$, measuring the fluid pressure required to deform the arch by an amount comparable to $d$. This motivates setting $P = p/p^{*}$. Inserting these scalings into the beam equation \eqref{eqn:beamdim}, we obtain 
\beq
\df{W}{X} +\tau^2\dd{W}{X} + P = 0, \qquad 0 < X < 1, \label{eqn:beamnondim}
\eeq
where $\tau^2 = T L^2/B$ is the dimensionless compressive force. From  \eqref{eqn:pressureSI}, the dimensionless pressure has the form
\beq
P(X) = P(1) + 12 Q \int_X^1 \frac{K(dW(\xi))}{\left[1-W(\xi)\right]^3}~\mathrm{d}\xi, \label{eqn:pressurenondim}
\eeq
where we have introduced the normalized flux
\beqn
Q  = \frac{\eta L^5 q_{\mathrm{in}}}{B b d^4}. 
\eeqn
This measures the ratio of the fluid pressure ($\sim \eta L q_{\mathrm{in}}/[b d^3]$) to the typical pressure required to deform the arch ($\sim p^{*}$). The pressure at the downstream clamp, \eqref{eqn:downstreampressureSI}, is written in dimensionless form as  
\beq
P(1) = 12 Q \frac{L_{\mathrm{d}}}{L}\frac{b}{b_{\mathrm{d}}} \left(\frac{d}{d_{\mathrm{d}}}\right)^3 \left[1-6 \left(\frac{2}{\pi}\right)^5 \frac{d_{\mathrm{d}}}{b_{\mathrm{d}}}\right]^{-1}. \label{eqn:downstreampressurenondim}
\eeq

From the expression \eqref{eqn:pressurenondim}, we see that $P  > 0$ for any physical displacement $W < 1$, i.e.~the viscous fluid always acts to oppose the displacement of the arch. (Note that because we are controlling the flux, the pressure $P$ appears to diverge as $W \nearrow 1$ to ensure that fluid can still be pushed through the channel.) The imposed end-shortening \eqref{eqn:constraintdim} becomes
\beq
\int_0^1\left(\d{W}{X}\right)^2~\id X=2 \frac{L \Delta L}{d^2} = \frac{\pi^2}{2}W_0^2,   \label{eqn:constraintnondim}
\eeq
where the last equality comes from solving for the buckled shape in the absence of any flow, $Q = 0$: this gives $W = W_0 (1-\cos 2\pi X)/2$  in terms of the channel blocking parameter $W_0 = w_0/d$. Finally, the clamped boundary conditions \eqref{eqn:clampbcsdim} are written as 
\beq
W(0) = W'(0) = W(1) = W'(1) = 0. \label{eqn:clampbcsnondim}
\eeq

Equations \eqref{eqn:beamnondim}--\eqref{eqn:pressurenondim} with constraints \eqref{eqn:constraintnondim}--\eqref{eqn:clampbcsnondim}  provide a closed system to determine the profile $W(X)$ and compressive force $\tau$. In a snapping experiment, the channel blocking parameter $W_0$ is fixed while we treat the normalized flux $Q$ as a control parameter that is quasi-statically varied. The other dimensionless parameters appearing in \eqref{eqn:pressurenondim} and \eqref{eqn:downstreampressurenondim} depend only on the geometry of the channel and are held constant throughout our experiments.  Their inclusion does not change the qualitative behavior of the system, so we do not consider their effect here.

\section{Equilibrium shapes}
\label{sec:equilibrium}
We now explore the equilibrium shapes of the arch as the flux $Q$ is varied. The cubic nonlinearity appearing in \eqref{eqn:pressurenondim} means we cannot make analytical progress in general; for each value of $W_0$ we instead solve the system \eqref{eqn:beamnondim}--\eqref{eqn:pressurenondim} with constraints \eqref{eqn:constraintnondim}--\eqref{eqn:clampbcsnondim} numerically in \textsc{matlab} using the routine \texttt{bvp4c}. Rather than performing continuation via the parameter $Q$, we instead control the compressive force $\tau$ and solve for the corresponding value of $Q$ at each stage. This allows us to avoid convergence issues near the snap-through bifurcation, and use a simple continuation algorithm that tracks equilibrium branches as $\tau$ is increased in small steps. At each stage, the solution at the previous value of $\tau$ is used as an initial guess for the update. For each equilibrium branch, the first guess is simply the Euler-buckling solution in the absence of any fluid flow, which is known analytically. 

When plotted in terms of $Q$, the resulting bifurcation diagram confirms that for small fluxes, both the `constricting' equilibrium shape (directed into the channel) and the `unconstricting' shape (directed away from the channel) exist. Moreover, a linear stability analysis confirms that these modes are linearly stable. However,  at larger fluxes this symmetry is broken, and the constricting shape eventually disappears at a saddle-node (fold) bifurcation when $Q = Q_{\mathrm{snap}}$: no constricting equilibrium exists for $Q > Q_{\mathrm{snap}}$. Any further increase in the fluid flux therefore causes the strip to snap to the unconstricting shape, which continues to exist and remain stable. This situation is in contrast to other snap-through instabilities in which the equilibrium becomes unstable rather than ceasing to exist \cite{Pandey2014SI}. 

\begin{figure}[h!]
\centering
\includegraphics[width=\textwidth]{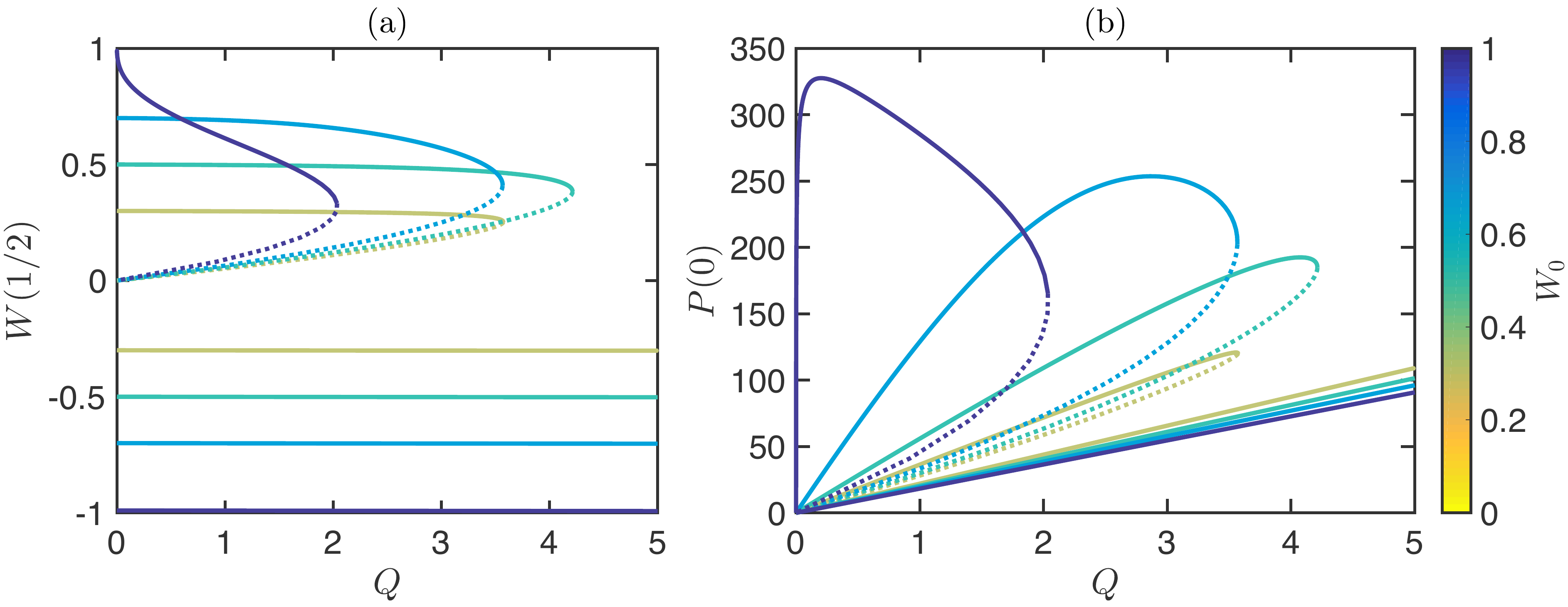} 
\caption{The equilibrium behavior of an elastic arch buckled into a lubrication-type flow. (a) Dimensionless midpoint displacement, $W(1/2) = w(L/2)/d$, as a function of the normalized upstream flux, $Q = \eta L q_{\mathrm{in}}/(B b d^4)$. Data is shown for $W_0 = w_0/d \in \left\lbrace 0.3,0.5,0.7,0.99\right\rbrace$ (indicated by the colorbar). The lower branches with $W(1/2) < 0$ correspond to the unconstricting or `snapped' shape. The upper branches correspond to the constricting shape, which disappears at a saddle-node bifurcation. (Note that there is a third, unstable branch, indicated by dotted curves.) (b) The corresponding pressure at the upstream end of the arch, $P(0) = p(0)/p^{*}$.}
\label{fig:response}
\end{figure}

Figure \ref{fig:response}a shows the bifurcation diagram for different values of the channel blocking parameter $W_0$; here we plot equilibrium modes in terms of their midpoint displacement $W(1/2)$ as a function of $Q$ (for the other dimensionless parameters, we use the values corresponding to our experimental system). This shows how the constricting shape (the upper, solid curves in figure \ref{fig:response}a) merges with an unstable mode (dashed curves) at the fold point; meanwhile the displacement in the unconstricting shape is relatively constant.

\begin{figure}
\centering
\includegraphics[width=\textwidth]{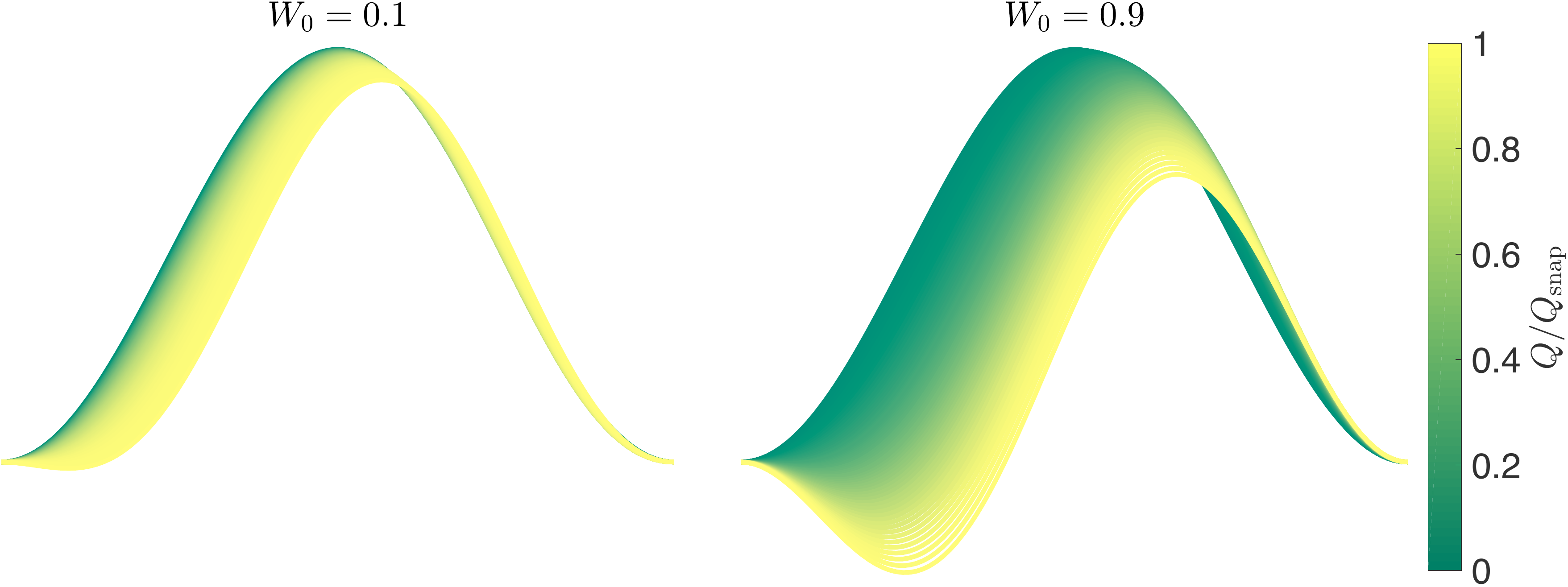} 
\caption{Sequence of equilibrium shapes prior to snapping. To facilitate comparison with different values of $W_0$, the vertical axes have been scaled to match the initial midpoint heights.}
\label{fig:shapes}
\end{figure}

The corresponding pressure at the upstream end of the arch, $P(0)$, is shown in figure \ref{fig:response}b. Comparing figures \ref{fig:response}a,b, we observe different regimes depending on the size of the channel blocking parameter, $W_0$. For small $W_0$, corresponding to shallow arch shapes, the channel is relatively unconstricted by the arch. This means that at low fluxes, the driving pressure needed is not sufficient to deform the arch: its midpoint displacement remains relatively constant. This is also evident in figure \ref{fig:shapes}, which plots the evolution of the arch shape for increasing flux when $W_0 = 0.1$; the shape of arch only changes very close to the snap-through transition. Away from snapping, the channel therefore acts as a rigid channel in this regime, with an effective conductivity that differs only slightly compared to a purely flat wall ($W = 0$). As the initial height $W_0$ increases, the critical flux needed for snap-through increases. 

For $W_0 \gtrsim 0.5$, the shape of the arch undergoes much larger changes prior to snapping (figure \ref{fig:response}a). As the channel becomes more constricted by the arch, a given flux creates a much larger driving pressure. This pressure quickly becomes sufficient to deform the arch, and is effective at `pushing' the midpoint away from the channel, so that the shape becomes increasingly asymmetric; this is evident in the sequence of shapes shown in figure \ref{fig:shapes} for $W_0 = 0.9$, and in Supplementary Movie 2. As the maximum height of the arch decreases, the width of the channel increases, which in turn lowers the driving pressure, even though the flux is still increasing. This underlies the non-monotonic pressure-flux relationship observed in this regime (figure \ref{fig:response}b) and also in our experiments. 

While arches with clamped ends exhibit snap-through under a variety of loading types, including point indentation and uniform pressure, we emphasize that the flow-induced snap-through here shows very different behavior. For example, in the case of point indentation, the arch always snaps at an Euler-buckling mode \citep{Pandey2014SI} as indentation proceeds, regardless of the arch height. This is in direct contrast to figure \ref{fig:shapes}, which shows that the sequence of shapes that the arch passes through before snapping depends strongly on $W_0$. Fundamentally, this dependence stems from the coupling between the arch elasticity and hydrodynamic pressure, as expressed by equation \eqref{eqn:pressurenondim}. As the arch comes close to blocking the channel ($W \nearrow 1$) this coupling becomes highly nonlinear and produces unexpected behavior that cannot be inferred by analogy with simpler loading types.

\subsection{Very shallow arches: $W_0 \ll 1$}
We can make analytical progress in the regime of very shallow arch shapes, $W_0 \ll 1$. To leading order, the arch is simply loaded by the linear pressure profile independent of the arch shape
\beqn
P(X) \sim P(1) + 12 K_0 Q (1-X),
\eeqn 
writing $K_0 = K(W = 0) = (1-6 \left[2/\pi\right]^5 d/b)^{-1}$. Using \eqref{eqn:downstreampressurenondim}, this can be written as
\beqn
P(X) \sim 12 K_0 Q(1+\epsilon_d -X),
\eeqn
where we introduce
\beqn
\epsilon_d = \frac{L_{\mathrm{d}}}{L}\frac{b}{b_{\mathrm{d}}}\left(\frac{d}{d_{\mathrm{d}}}\right)^3 \frac{1-6 \left(2/\pi\right)^5 d/b}{1-6 \left(2/\pi\right)^5 d_{\mathrm{d}}/b_{\mathrm{d}}}.
\eeqn
This parameter compares the conductivity of the flexible part of the channel (with $W_0\ll1$) to the conductivity of the channel further downstream. The corresponding solution of the beam equation \eqref{eqn:beamnondim} satisfying the clamped boundary conditions \eqref{eqn:clampbcsnondim} is
\beq
W(X) = K_0 Q \left\lbrace \frac{2}{\tau^2}\left[X^3 -3(1+\epsilon_{\mathrm{d}})X^2\right] + C\left(\cos\tau X - 1\right) + D\left(\sin\tau X-\tau X\right)\right\rbrace, \label{eqn:steadyWapprox}
\eeq 
where 
\begin{eqnarray*}
C & = & \frac{(2+3\epsilon_{\mathrm{d}})\tau(\cos\tau-1)+3(1+2\epsilon_{\mathrm{d}})(\tau-\sin\tau)}{\tau^3\sin\tau[\tan(\tau/2)-\tau/2]}, \\
D & = & \frac{(2+3\epsilon_{\mathrm{d}})\tau\sin\tau+3(1+2\epsilon_{\mathrm{d}})(\cos\tau-1)}{\tau^3\sin\tau[\tan(\tau/2)-\tau/2]}.
\end{eqnarray*}
To determine $\tau$ in terms of the control parameter $Q$, we substitute \eqref{eqn:steadyWapprox} into the end-shortening constraint \eqref{eqn:constraintnondim}. Because the term in braces in \eqref{eqn:steadyWapprox} is independent of $Q$ (note $C$ and $D$ depend only on $\tau$ and $\epsilon_{\mathrm{d}}$), carrying out the integration leads to an implicit equation of the form
\beq 
\left(\frac{Q}{W_0}\right)^2 = f(\tau).\label{eqn:tauimplicit}
\eeq
The function $f(\tau)$ depends only on the geometry of the channel (via $K_0$ and $\epsilon_{\mathrm{d}}$) and may be written in closed form, though as the expression is rather long we do not present it here. 

For each value of $Q/W_0$, this relation may be used to determine the possible values of $\tau$ numerically. The corresponding midpoint displacement is then evaluated using \eqref{eqn:steadyWapprox} as
\beqn
W(1/2) = 3 K_0 Q(1+2\epsilon_{\mathrm{d}})\frac{\tau/4-\tan(\tau/4)}{\tau^3}.
\eeqn
In particular, we are able to find the value of  $Q/W_0$ at the saddle-node bifurcation. This yields the following scaling law for the critical flux required for snap-through in this regime:
\beq 
Q_{\mathrm{snap}} \propto W_0  \quad \mathrm{when} \quad W_0 \ll 1,
\eeq
where the pre-factor depends only on the channel geometry. For our experimental system ($ d = 6~\mathrm{mm}$, $b = 23~\mathrm{mm}$, $L = 50~\mathrm{mm}$, $d_\mathrm{d} = 5~\mathrm{mm}$, $b_\mathrm{d} = 21~\mathrm{mm}$ and $L_\mathrm{d} = 22.2~\mathrm{mm}$) we compute 
\beqn
\epsilon_{\mathrm{d}} \approx 0.8262, \quad K_0 \approx 1.196, \quad Q_{\mathrm{snap}} \approx 16.04 W_0, \quad \tau_{\mathrm{snap}} \approx 8.597,  \quad W_{\mathrm{snap}}(1/2) \approx 0.8841 W_0. 
\eeqn 
The prediction $Q_{\mathrm{snap}} \approx 16.04 W_0$ is plotted as a black dashed line in figure $3$ of the main text. The fact that the pre-factor in  $W_{\mathrm{snap}}(1/2)$ is close to unity explains the small change in arch shape prior to snapping that is observed in this regime (figure \ref{fig:shapes}). 

\section{Reversible snap-through}
\label{sec:reversible}
In the  setup illustrated in figure 1a of the main text, the arch is bistable in the absence of any flow: both constricting and unconstricting equilibrium shapes exist and are stable. This is due to the up-down symmetry of the horizontally clamped arch in the absence of flow: both states are simply reflections of one another. As a consequence, once flow causes the arch to snap to the unconstricting configuration, it remains in this state when the flow rate is subsequently reduced --- the snap is irreversible. This history dependence may be useful in some applications, for example in detecting whether a given flow rate has been exceeded. In other scenarios, however, it may instead be desirable to have a reversible snap-through so that, for example, the arch returns to its original state when the flow is switched off, without the need for direct intervention.

To obtain such a reversible snap-through, we must break the up-down symmetry so that only the constricting state is stable in the absence of any flow. This may be achieved in a variety of ways, including introducing external constraints that limit the arch displacement, or by using a restoring force to `push' the arch back to its original position when it exceeds the fluid pressure. However, a simpler alternative is to vary the boundary conditions that are applied at the ends of the arch. 

\begin{figure}
\centering
\includegraphics[scale = 1.4]{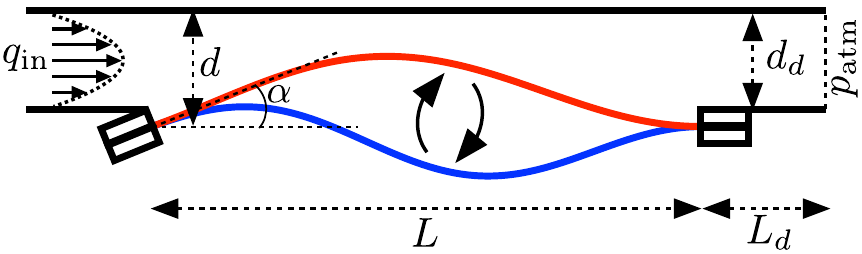} 
\caption{Modifying the boundary conditions applied to the arch, e.g.~by clamping the upstream end at an angle $\alpha > 0$, allow a reversible snap-through to be obtained.}
\label{fig:reversibleschematic}
\end{figure}

In this section we focus on perhaps the simplest possible modification: the upstream end of the arch is  clamped at an angle $\alpha > 0$ to the flow direction (rather than parallel to the flow direction), while the other end remains clamped parallel to the flow direction a distance $L < L_{\mathrm{strip}}$ further downstream. This is illustrated schematically in figure \ref{fig:reversibleschematic}. The behaviour of such a tilted arch  has been studied in the absence of flow by \cite{Gomez2017SI}. The key result is that for sufficiently large $\alpha$, no unconstricting state exists --- the only equilibrium solution is a constricting state as drawn in figure  \ref{fig:reversibleschematic} (red curve). This same phenomenology should also hold for our system when the fluid flux is sufficiently small. In an experiment, we therefore expect that the arch would still snap to an unconstricting state (blue curve) as the flow rate is increased, but that, as the flux is subsequently reduced back to zero, this state will eventually become unstable and snap back to the original shape.

Provided that $0 < \alpha \ll 1$, the shape of the arch will remain shallow, so we can use linear beam theory and lubrication theory as in \S\ref{sec:theory}. The dimensionless equations \eqref{eqn:beamnondim}--\eqref{eqn:pressurenondim} with imposed end-shortening \eqref{eqn:constraintnondim} then also apply. The only change is in the clamped boundary conditions \eqref{eqn:clampbcsnondim}, which become
\beq
W(0) = 0, \quad W'(0) = \frac{\alpha L}{d}, \quad W(1) = W'(1) = 0. \label{eqn:reversibleBC}
\eeq 
We also note that the last equality in \eqref{eqn:constraintnondim} is no longer valid: when $\alpha \neq 0$, the buckled shape in the absence of any flow differs from classic Euler-buckling, so that the relation between the channel blocking parameter $W_0$ and the end-shortening $\Delta L$ is no longer simple. We shall therefore refer to the value of $L\Delta L/d^2$ in this section, rather than referring to the channel blocking parameter $W_0$.

\subsection{Absence of flow}
\begin{figure}
\centering
\includegraphics[width= 0.7\textwidth]{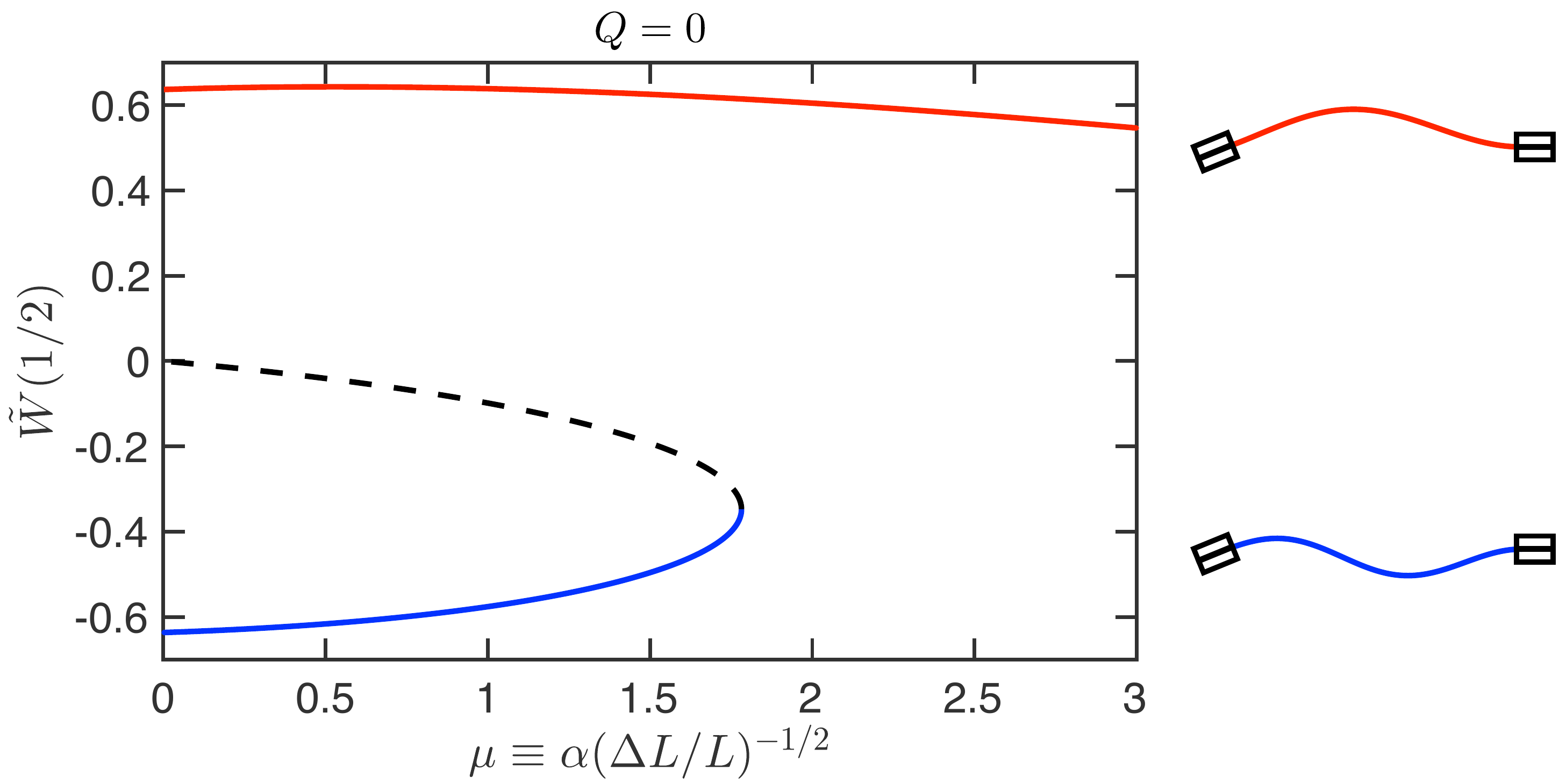} 
\caption{Response diagram in the absence of any flow $(Q=0)$ when the upstream end is clamped at an angle $\alpha > 0$.}
\label{fig:reversiblezeroflow}
\end{figure}

We briefly review the equilibrium states in the absence of any flow. When $Q = 0$, we have two parameters in the problem: these are the dimensionless end-shortening $L\Delta L/d^2$, and the normalized inclination angle $\alpha L/d$ appearing in \eqref{eqn:reversibleBC}. It is possible to scale out the end-shortening by setting
\beqn
W(X) = \left(\frac{L\Delta L}{d^2}\right)^{1/2}\tilde{W}(X).
\eeqn
The beam equation \eqref{eqn:beamnondim}, end-shortening constraint  \eqref{eqn:constraintnondim} and clamped boundary conditions \eqref{eqn:reversibleBC} then become
\begin{eqnarray*}
&& \df{\tilde{W}}{X} +\tau^2\dd{\tilde{W}}{X} = 0, \\
&& \int_0^1\left(\d{\tilde{W}}{X}\right)^2~\id X=2, \quad \tilde{W}(0) = 0, \quad \tilde{W}'(0) = \mu, \quad \tilde{W}(1) = \tilde{W}'(1) = 0.
\end{eqnarray*}
Here we have introduced the geometric parameter
\beqn
\mu = \alpha \left(\frac{\Delta L}{L}\right)^{-1/2},
\eeqn
which measures the ratio of the inclination angle $\alpha$ to the typical arch slope due to the imposed end-shortening, $(\Delta L/L)^{1/2}$. Using an analytical solution, it was shown in \cite{Gomez2017SI} that for $0 < \mu < \mu_{\mathrm{fold}} \approx 1.7818$, both constricting and unconstricting shapes exist and are stable. However, the unconstricting shape disappears at a saddle-node bifurcation when $\mu = \mu_{\mathrm{fold}}$ (it meets an unstable mode that is not observed experimentally), so that only the constricting equilibrium exists for $\mu > \mu_{\mathrm{fold}}$. This is illustrated in figure \ref{fig:reversiblezeroflow}, which plots the rescaled midpoint height $\tilde{W}(1/2)$ of the various modes as functions of $\mu$. 

\subsection{Equilibrium shapes for $Q > 0$}
Based on the above discussion, we expect that snap-through is reversible provided $\mu > \mu_{\mathrm{fold}}$. In practice, this can be satisfied simultaneously with the condition $0 < \alpha \ll 1$, needed for a shallow arch shape, by choosing $\Delta L/L$ sufficiently small. To confirm this intuition we now consider the equilibrium shapes as $Q$ is quasi-statically varied from zero.

We solve the system \eqref{eqn:beamnondim}--\eqref{eqn:pressurenondim} with constraint \eqref{eqn:constraintnondim} and boundary conditions  \eqref{eqn:reversibleBC} numerically in \textsc{matlab} using the routine \texttt{bvp4c}. We anticipate the bifurcation diagram to be more complex than the case $\alpha = 0$ (since, for example, we expect to observe an additional `snap-back' when the flow rate is subsequently reduced). Rather than using a simple continuation algorithm as in \S\ref{sec:equilibrium}, we therefore use a pseudo-arclength continuation algorithm  \cite{seydel2009}. This involves introducing an abstract parameter that paramaterizes arclength along equilibrium branches, and at each stage we solve for both the flux $Q$ and tension $\tau$ as part of the problem. To begin the continuation, we use analytical solutions for the constricting and unconstricting shapes in the absence of flow, $Q = 0$ \cite{Gomez2017SI}.

\begin{figure}
\centering
\includegraphics[width=0.8\textwidth]{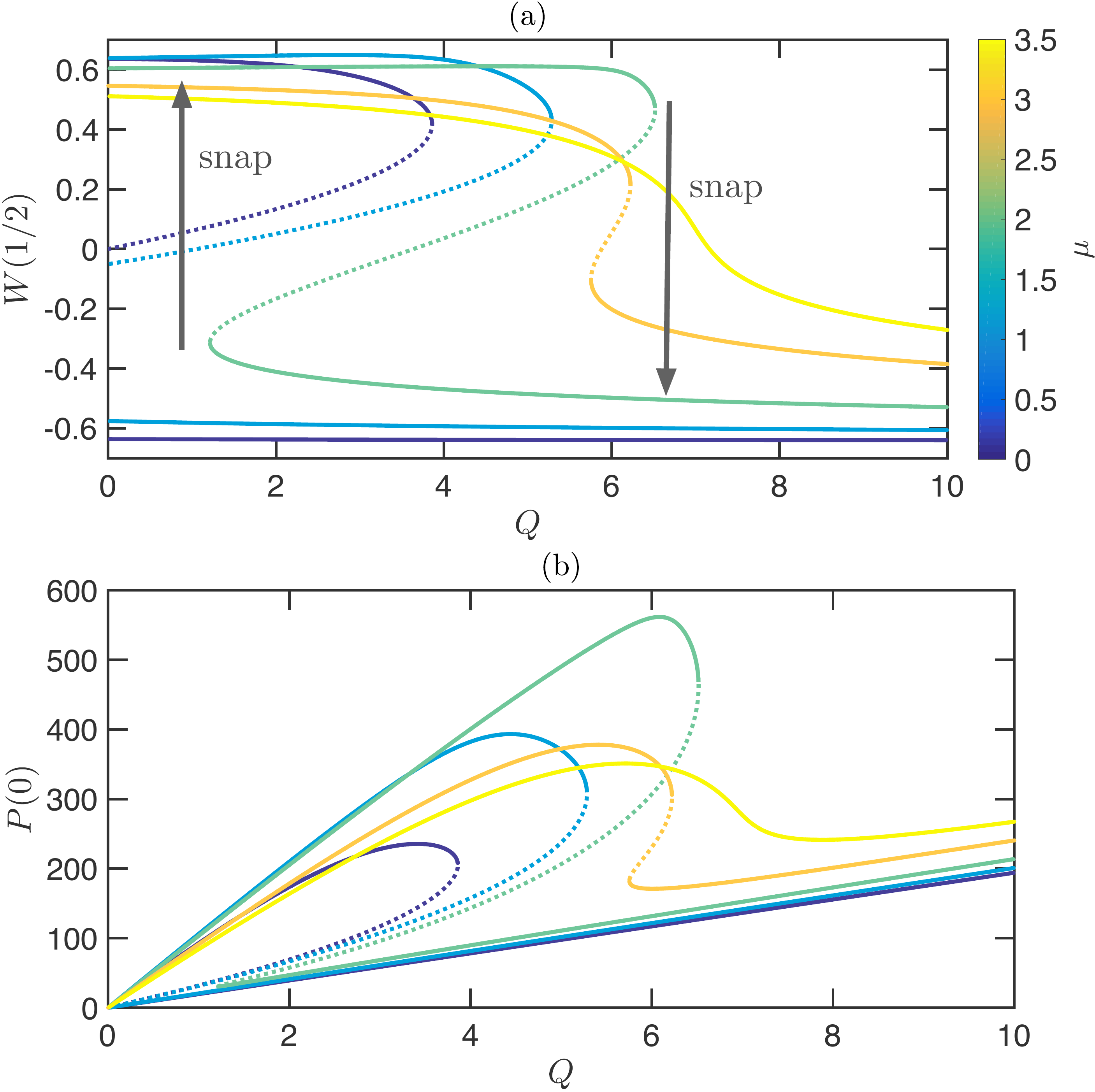}
\caption{Equilibrium behavior for the reversible snap-through setup (here $L\Delta L/d^2 = 1$, and all other parameters are the values corresponding to our experimental system). (a) Dimensionless midpoint displacement $W(1/2)$ as a function of the normalized flux $Q$. Data is shown for $\mu \in \left\lbrace 0,1,2,3,3.5\right\rbrace$ (indicated by the colorbar). The lower branches with $W(1/2) < 0$ correspond to the unconstricting state, while the upper branches correspond to the constricting shape (dotted curves indicate unstable branches.) (b) The corresponding pressure at the upstream end of the arch, $P(0)$.}
\label{fig:reversibleresponse}
\end{figure}

We find that when $\mu < \mu_{\mathrm{fold}}$, the bifurcation diagram is qualitatively similar to the case $\alpha = 0$: the unconstricting state exists for all fluxes $Q$, and the equilibrium branch is disconnected from the constricting equilibrium branch. The snap-through in this case is therefore irreversible, as might be expected. However, when  $\mu > \mu_{\mathrm{fold}}$, both branches are connected by a succession of two folds. This corresponds to a hysteresis loop: the arch snaps from the constricting state to the unconstricting state upon increasing the flux, and then snaps back (at a flux lower than the first snap) when the flux is subsequently reduced. At still larger $\mu$, these folds disappear and the solution instead smoothly transitions from a constricting to an unconstricting  state --- no snap-through occurs. This is shown in figure \ref{fig:reversibleresponse}a, which plots the midpoint displacement of equilibrium modes as a function of $Q$.  We see from the figure that reversible snap-through behavior is obtained in the range $\mu_{\mathrm{fold}} < \mu \lesssim 3$. The corresponding pressure at the upstream end of the arch is shown in figure \ref{fig:reversibleresponse}b.

In addition, we observe several interesting nonlinear features when $\alpha > 0$. In particular, there is a regime where the midpoint displacement initially increases slightly with the flux, despite the fact that the fluid pressure in the channel, which opposes the arch displacement, is increasing (figure \ref{fig:reversibleresponse}a). Surprisingly, we also see that at a given flux, the driving pressure $P(0)$ does not simply increase monotonically with $\mu$ (figure \ref{fig:reversibleresponse}b). This is due to the non-monotonic relationship between $W(1/2)$ and $\mu$ that is observed in figure \ref{fig:reversiblezeroflow}.

\section{A passive fluid `fuse'}
\label{sec:fuse}
\begin{figure}
\centering
\includegraphics[scale = 1.4]{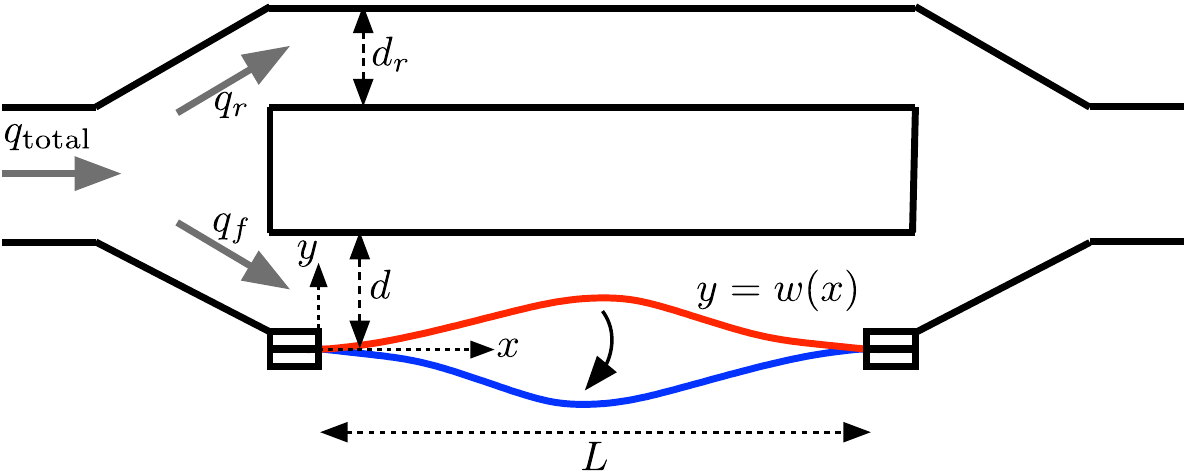} 
\caption{Schematic of a passive flow limiter, which places a rigid channel in parallel with a flexible channel.}
\label{fig:fuseschematic}
\end{figure}

In this section we discuss the scenario in which a channel containing an arch element is placed in parallel with a second, purely rigid, channel; see figure \ref{fig:fuseschematic}. For the flexible channel, we use the same notation as in \S\ref{sec:theory}: the channel has width $d$ and depth $b$ when the arch is flat, and the displacement of the arch is $y = w(x)$. The geometric properties of the rigid channel are its width $d_r$ and depth $b_r$. We again consider the case when the ends of the arch are clamped parallel to the flow direction a distance $L$ apart, and the midpoint height is $w_0$ in the absence of any flow. The volumetric flux through the flexible and rigid channels are $q_f$ and $q_r$ respectively, with the total flux $q_{\mathrm{total}} = q_r + q_f$. 

In the lubrication approximation, the pressure gradient in the flexible channel is given by the Poiseuille law:
\beq
-\frac{b[d-w(x)]^3}{12 \eta} \frac{\mathrm{d} p}{\mathrm{d}x} = q_f, \quad 0 < x < L,  \label{eqn:flexiblegradient}
\eeq
where $\eta$ is the dynamic viscosity of the working fluid, and we neglect finite-depth effects for simplicity (i.e.~we set $K(w) \equiv 1$). In the rigid channel, we instead have 
\beq
-\frac{b_r d_r^3}{12 \eta} \frac{\mathrm{d} p}{\mathrm{d}x} = q_r. \label{eqn:rigidgradient}
\eeq
Because the channels are in parallel, they are both subject to the same pressure drop $\Delta p$. We assume that the pressure gradient outside of the interval $(0,L)$ is negligible in both channels, i.e.~the flow resistance is dominated by the interval containing the arch. This is roughly the situation drawn in figure \ref{fig:fuseschematic}. Integrating \eqref{eqn:flexiblegradient} and \eqref{eqn:rigidgradient}, we obtain
\beq
\Delta  p = \frac{q_f}{k_f} = \frac{q_r}{k_r}, \label{eqn:deltap}
\eeq
where we have defined the effective hydraulic conductivities  
\beqn
k_f = \frac{b}{12\eta \int_0^{L} \left[d-w(\xi)\right]^{-3}\id\xi}, \qquad k_r = \frac{b_r d_r^3}{12\eta L}.
\eeqn
Note that $k_f$ depends on the flux $q_f$, since the arch shape depends on the fluid loading in the flexible channel.

\subsection{Non-dimensionalization}
We introduce dimensionless variables based on the geometry of the flexible channel (as in \S\ref{sec:theory}), setting
\beqn
X = \frac{x}{L}, \quad W =  \frac{w}{d}, \quad W_0 =  \frac{w_0}{d}, \quad P = \frac{p}{Bd/L^4}, \quad (Q_{\mathrm{total}},Q_r,Q_f) =  \frac{\eta L^5}{Bb  d^4}(q_{\mathrm{total}},q_r,q_f).
\eeqn
Under these rescalings, we can combine \eqref{eqn:deltap} with $Q_{\mathrm{total}} = Q_r + Q_f$ to obtain implicit equations for $Q_f$ of the form
\begin{eqnarray}
Q_{\mathrm{total}} & = & Q_f \left(1+\frac{k_r}{k_f}\right),  \label{eqn:Qtotal} \\
Q_r & = & Q_f \frac{k_r}{k_f}. \label{eqn:Qr}
\end{eqnarray}
The ratio of conductivities, $k_r/k_f$, depends on its value for a flat arch, $\lambda = b_r d_r^3/(b d^3)$, and the change due to the dimensionless displacement $W(X)$. We find that
\beq
\frac{k_r}{k_f}  = \lambda I, \label{eqn:conductratio}
\eeq 
where we define
\beqn
I =  \int_0^1 \frac{\id\xi}{\left[1 - W(\xi) \right]^3}.
\eeqn
The dimensionless arch shape $W(X)$ satisfies the system \eqref{eqn:beamnondim}--\eqref{eqn:pressurenondim} with constraints \eqref{eqn:constraintnondim}--\eqref{eqn:clampbcsnondim} that were derived in \S\ref{sec:theory}, though now we set $Q \to Q_f$, $P(1) = 0$ and $K(w) = 1$. 

In an experimental system, the relevant control parameter is the total flux $Q_{\mathrm{total}}$. However, it is mathematically convenient to instead control $Q_f$. For each channel blocking parameter $W_0$, we then solve the system \eqref{eqn:beamnondim}--\eqref{eqn:pressurenondim} with constraints \eqref{eqn:constraintnondim}--\eqref{eqn:clampbcsnondim} numerically in \textsc{matlab} using the routine \texttt{bvp4c}. This determines the arch shape $W(X)$. At each step in the continuation, we use quadrature to evaluate the integral $I$ appearing in  the conductivity ratio \eqref{eqn:conductratio}.  This produces pairs of values $(Q_f,I)$ for each equilibrium shape. Note that, because we are controlling the flux in the flexible channel, this procedure can be performed independently of $\lambda$. For a specified $\lambda$, we can then use equations \eqref{eqn:Qtotal}--\eqref{eqn:Qr} to determine the corresponding pairs of values of $(Q_r,Q_{\mathrm{total}})$ as $Q_f$ is varied. In this way, we construct a parametric plot of $q_r/q_{\mathrm{total}}$ as a function of $Q_{\mathrm{total}}$.

Numerical results for the case $W_0 = 0.99$ are given in figure $4$ of the main text. (Here the constricting shape is only plotted until snapping occurs under changes in $Q_{\mathrm{total}}$, as would occur experimentally.) We see that if $\lambda$ is very small, the deformation of the arch prior to snapping has a large influence on the system, causing a significant decrease in $q_r/\qtotal$ before the critical flux is reached. Hence there is a trade-off: for small $\lambda$ the total change in $q_r/\qtotal$ is large but some of this change occurs smoothly before short-circuiting, while for larger $\lambda$ the drop in  $q_r/\qtotal$ is almost entirely due to snapping, but is of a smaller degree.

\subsection{Analytical results for fuse-like behavior}
To gain further insight, we consider the fraction of the total flux that passes through the rigid channel. From equations \eqref{eqn:Qtotal}--\eqref{eqn:conductratio}, this is given by
\beq
\frac{q_r}{q_{\mathrm{total}}} = \frac{Q_r}{Q_{\mathrm{total}}} = \frac{\lambda I}{1 + \lambda I}. \label{eqn:fluxratio}
\eeq
We note that $\lambda$ is fixed once the geometry of each channel is specified, while $I$ depends on the arch shape and so will change with the flux.  In particular, there are three values of $I$ that characterize the properties and effectiveness of the fuse: the value in the unconstricting shape, denoted $I_\text{uc}$, the value in the initial constricting shape (i.e.~with no fluid flow), denoted $I_\text{c}$, and the value in the constricting shape at the point of snapping, denoted $I_\text{snap}$.  

When there is no flux, the constricting shape satisfies $W(X) = W_0 (1-\cos 2\pi X)/2$ (this is the Euler-buckling solution), and so by direct integration we have
\beq
I_\text{c} = \frac{8-W_0 (8-3 W_0)}{8(1-W_0)^{5/2}}.  \label{eqn:Iconstrict}
\eeq
In the unconstricting configuration, the shape of the arch always remains close to the buckled shape in the absence of any flow (see the lower branches in figure \ref{fig:response}a). This gives $W(X) \approx -W_0 (1-\cos 2\pi X)/2$, for which we can compute
\beq
I_\text{uc} = \frac{8+W_0 (8+3 W_0)}{8(1+W_0)^{5/2}}.  \label{eqn:Iunconstrict}
\eeq
Considering the behavior of the right-hand-side as $W_0$ varies over the interval $(0,1)$, we find that $I_\text{uc} \in (0.4,1)$, so in particular $I_\text{uc}=O(1)$ for any arch. From  \eqref{eqn:fluxratio}, to ensure that $q_r/q_{\mathrm{total}} \ll 1$ after snapping occurs (i.e.~effective `short-circuiting'), we therefore need to choose $\lambda \ll1$; this corresponds to a much larger conductivity in the flexible channel compared to the rigid channel when the arch is flat.

As well as directing most of the fluid to the flexible channel after snapping, an effective fuse should have most of the fluid passing through the rigid channel prior to snapping (i.e.~$q_r/q_{\mathrm{total}} \approx 1$). This ensures a switch-like response with little other disruption to the flow through the rigid channel.   From \eqref{eqn:fluxratio}, this requires $\lambda I_\text{c} \gg 1$ and $\lambda I_\text{snap} \gg 1$. To be consistent with the geometrical constraint $\lambda \ll 1$  above, this suggests we need $I_\text{c} \gg 1/\lambda \gg 1$ and $I_\text{snap} \gg 1/\lambda \gg 1$.  

Considering \eqref{eqn:Iconstrict}, we see that $I_\text{c} \gg 1$ requires the arch to block most of the flexible channel, $W_0 \approx 1$. However, when $W_0 \approx 1$, the constricting shape also undergoes a large shape change prior to snapping (see figures \ref{fig:response}a and \ref{fig:shapes}), meaning that $I_\text{snap}$ can be significantly smaller from $I_\text{c}$.  We can estimate this decrease by numerically computing $I_\text{snap}$. In general, this depends on the properties of the rigid channel (via $\lambda$) so that analytical progress is not possible. Moreover, the calculation is complicated by the fact that experimentally we do not control the flux through the flexible channel, as in \S\ref{sec:theory}, but rather the total flux $Q_{\mathrm{total}}$. However, we can obtain a reasonable estimate for $I_\text{snap}$ by using the shape at the fold bifurcation for a single flexible channel (in which case the critical flux is independent of $\lambda$). For example, in the case of $W_0=0.99$ we compute $I_\text{snap} \approx 4.3$ while $I_\text{c} \approx 37.8\times 10^4$ and $I_\text{uc} \approx 0.42$.

\begin{figure}
\centering
\includegraphics[width= 0.6\textwidth]{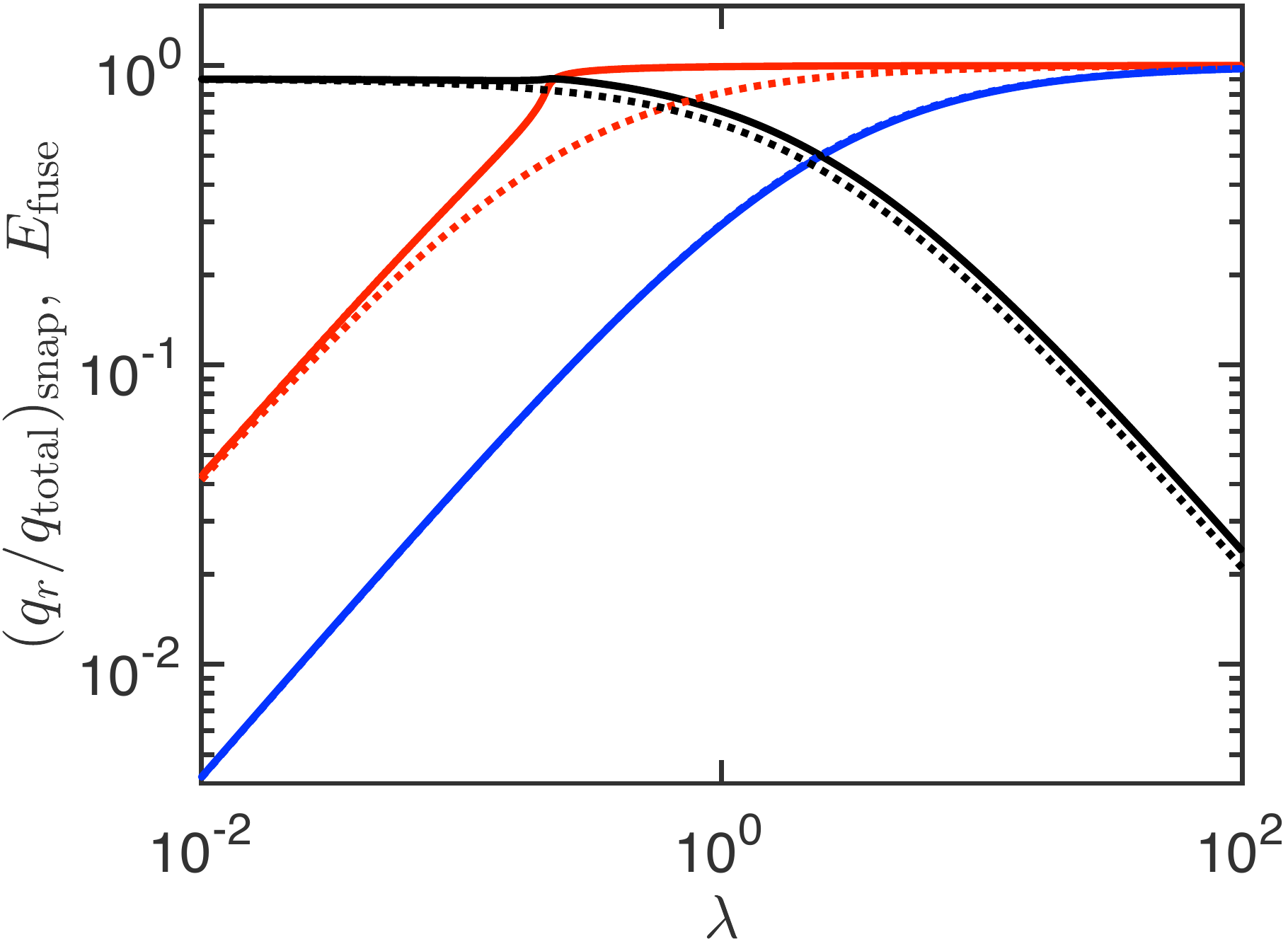} 
\caption{Proportion of the total flux received by the rigid channel at the snapping transition,  $(q_r/\qtotal)_{\mathrm{snap}}$, as a function of the conductivity ratio $\lambda$ ($W_0 = 0.99)$. Red/blue curves correspond to the constricting/unconstricting shape, while the black curve gives the fuse efficiency $E_{\mathrm{fuse}}$ defined in \eqref{eqn:Efuse}. Both numerical results (solid curves) and the approximations \eqref{eqn:fluxratiopresnap}--\eqref{eqn:Efuse} (dotted curves) are shown.}
\label{fig:fuseefficiency}
\end{figure}

In this way,  the fraction of the total flux through the rigid channel just before snapping can be estimated using  \eqref{eqn:fluxratio} as
\beq
\left(\frac{q_r}{q_{\mathrm{total}}}\right)_{\mathrm{snap,~constricting}} \approx   \frac{\lambda}{\lambda+\alpha(W_0)}, \label{eqn:fluxratiopresnap}
\eeq
where $\alpha(W_0) = 1/I_\text{snap}(W_0)$. Returning to the result \eqref{eqn:Iunconstrict} for the unconstricting shape, and expanding for $W_0 \approx 1$, the fraction of the total flux through the rigid channel immediately after snapping is then
\beq
\left(\frac{q_r}{q_{\mathrm{total}}}\right)_{\mathrm{snap,~unconstricting}} \approx   \frac{\lambda}{\lambda+32\sqrt{2}/19}. \label{eqn:fluxratiopostsnap}
\eeq
The efficiency of the fuse,  defined as the decrease in $q_r$ caused by snap-through divided by its value prior to snap-through, is 
\beq
E_{\mathrm{fuse}} = 1- \frac{\left(q_r/q_{\mathrm{total}}\right)_{\mathrm{snap,~unconstricting}} }{\left(q_r/q_{\mathrm{total}}\right)_{\mathrm{snap,~constricting}} } \approx \frac{32\sqrt{2}/19-\alpha(W_0)}{\lambda+32\sqrt{2}/19}. \label{eqn:Efuse}
\eeq

The expressions \eqref{eqn:fluxratiopresnap}--\eqref{eqn:Efuse}  show that small $\lambda$ leads to a greater fuse efficiency, but at the expense of a smaller flux through the rigid channel before snapping occurs. These expressions are plotted in figure \ref{fig:fuseefficiency} (dotted curves) for the case $W_0 = 0.99$, when we compute $\alpha(W_0)\approx 0.23$. For comparison we also plot  the numerically determined values of $(q_r/\qtotal)_{\mathrm{snap}}$ and $E_{\mathrm{fuse}}$ (solid curves). We see that the analytical predictions provide a good approximation of the numerical solution over a large range of $\lambda$. The numerics also show a rapid decrease in $(q_r/\qtotal)_{\mathrm{snap}}$ in the constricting shape for moderately small $\lambda$  ($\lambda\approx 0.2$), where the efficiency $E_{\mathrm{fuse}}$ attains a global maximum ($E_{\mathrm{fuse}} \approx 0.9$); at smaller values of $\lambda$ the efficiency remains relatively constant. In applications, an optimum choice of $\lambda$ might therefore be the value at which this maximum efficiency is attained.

Finally, we also note that after snap-through occurs, the hysteresis we observe in the snapped state ensures the rigid channel will retain the history of the short-circuiting if the flux is subsequently reduced.

\section{Scalability}\label{sec:scale}
We consider the scalability of elastic snap-through to smaller, microfluidic, devices. A typical microfluidic channel may have length $L \sim 1~\mathrm{mm}$, depth $d \sim 200~\mu\mathrm{m}$, and width $b \sim 500~\mu\mathrm{m}$. Assuming that the liquid used is water ($\eta=10^{-3}\mathrm{~Pa}~\mathrm{s}$) we find that the dimensional critical flux at which snap-through occurs is $\qfold\approx B/\mathrm{Pa\,s}$: with a flexible element of PDMS ($E \sim 1~\mathrm{MPa}$) and thickness in the range $10\mathrm{~\mu m}\lesssim h\lesssim 100\mathrm{~\mu m}$ we find $6\mathrm{~\mu L}~\mathrm{min}^{-1}\lesssim\qfold\lesssim6\mathrm{~mL}~\mathrm{min}^{-1}$. These ranges of snap-through fluxes are well within experimentally obtainable fluxes, which are typically \cite{stone2004SI,oh2006SI} in the range $ 10\mathrm{~\mu L}~\mathrm{min}^{-1} \lesssim \qin\lesssim10\mathrm{~mL}~\mathrm{min}^{-1}$. We also note that the relative sensitivity of the critical flux $\qfold\sim Bbd^4/(\eta L^5)$ to the channel geometry means that this critical flux may very easily be tuned to a desired value outside the range discussed above. 
 
\section{Supplementary Movies}
\label{sec:movie}
\setcounter{figure}{0}   
\renewcommand{\figurename}{Movie}

\begin{figure}[h!]
\centering
\includegraphics[width = 0.6\textwidth]{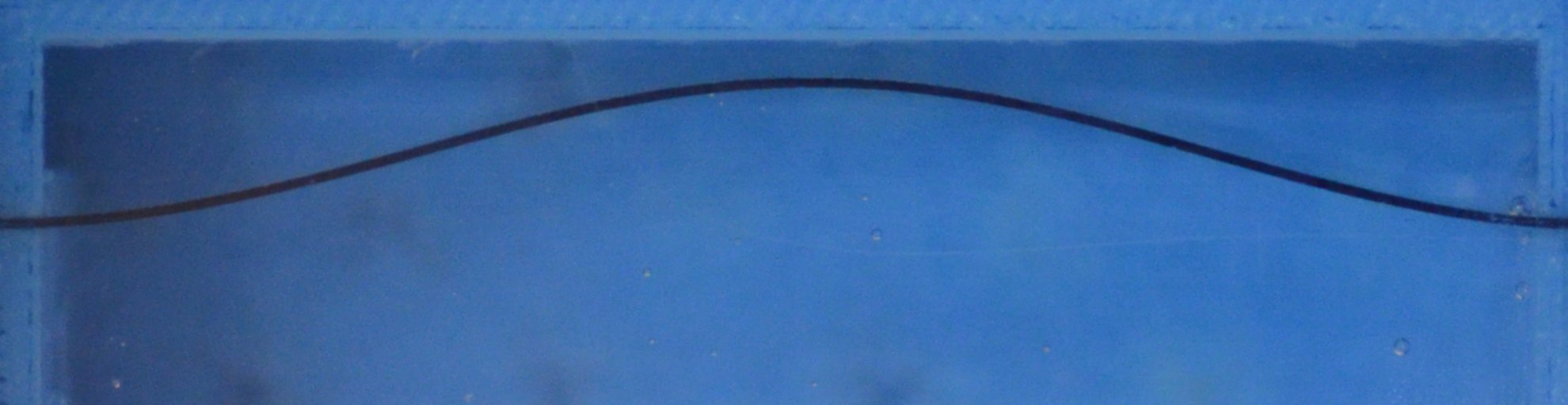} 
\caption{Top view showing the evolution of the arch shape during a snapping experiment ($h = 0.25~\mathrm{mm}$, $w_0 = 4.7~\mathrm{mm}$, $\eta = 1.60 \pm 0.10~\mathrm{Pa}~\mathrm{s}$). The arch is marked black along its edge with its depth into the page. The volumetric flux is ramped linearly from zero at a rate $70~\mathrm{mL}~\mathrm{min}^{-2}$. Flow occurs from left to right in the part of the channel above the arch; below the arch is a stationary bath of fluid. The movie was recorded at $1$ frame per second, and plays back at $5$ frames per second. }
\end{figure}

\begin{figure}[h!]
\centering
\includegraphics[width = 0.6\textwidth]{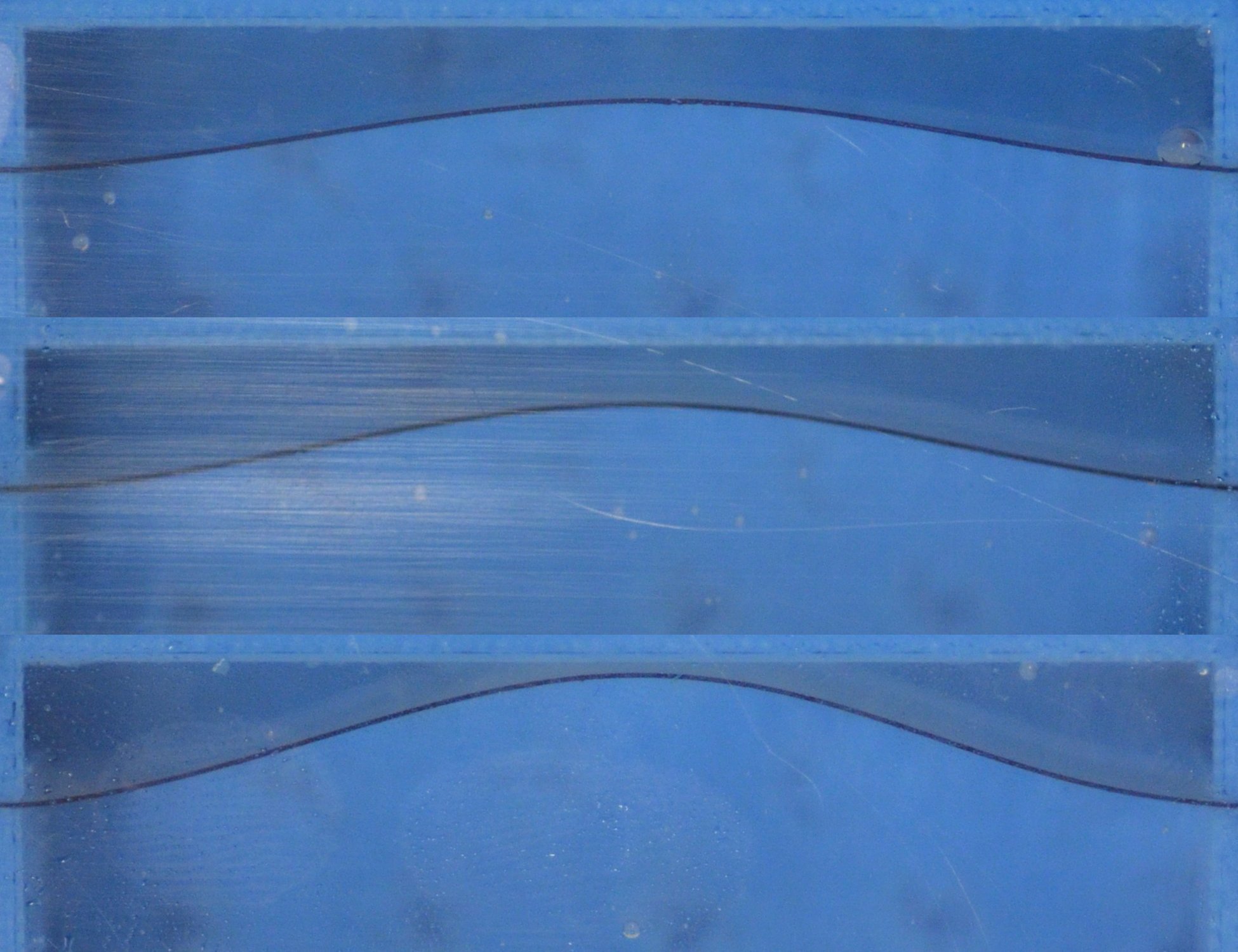} 
\caption{The evolution of the arch shape for different initial midpoint heights $w_0$ ($h = 0.1~\mathrm{mm}$, $\eta = 1.20 \pm 0.10~\mathrm{Pa}~\mathrm{s}$, ramping rate $2~\mathrm{mL}~\mathrm{min}^{-2}$). Here $w_0 = 2.9~\mathrm{mm}$ (top), $w_0 = 3.6~\mathrm{mm}$ (middle) and $w_0 = 5.5~\mathrm{mm}$ (bottom). This shows how for large arch heights ($w_0/d \gtrsim 0.5)$, the shape of the arch undergoes much larger changes prior to snapping, with the critical flux decreasing as $w_0$ increases. Each movie was recorded at $1$ frame every $2$ seconds, and plays back at $10$ frames per second.}
\end{figure}


\begin{thebibliography}{10}

\bibitem{Khoo:2011wp}
C.~K. Khoo, F.~Salim, and J.~Burry,
\newblock Int. J. Arch. Comput. {\bf 9}, 397 (2011).

\bibitem{stone2004}
H.~A. Stone, A.~D. Stroock, and A.~Ajdari,
\newblock Annu. Rev. Fluid Mech. {\bf 36}, 381 (2004).

\bibitem{unger2000}
M.~A. Unger, H.-P. Chou, T.~Thorsen, A.~Scherer, and S.~R. Quake,
\newblock Science {\bf 288}, 113 (2000).

\bibitem{oh2006}
K.~W. Oh and C.~H. Ahn,
\newblock J. Micromech. Microengng {\bf 16}, R13 (2006).

\bibitem{holmes2013}
D.~P. Holmes, B.~Tavakol, G.~Froehlicher, and H.~A. Stone,
\newblock Soft Matter {\bf 9}, 7049 (2013).

\bibitem{tavakol2014}
B.~Tavakol, M.~Bozlar, C.~Punckt, G.~Froehlicher, H.~A. Stone, I.~A. Aksay, and
  D.~P. Holmes,
\newblock Soft Matter {\bf 10}, 4789 (2014).

\bibitem{weibel2005}
D.~B. Weibel, M.~Kruithof, S.~Potenta, S.~K. Sia, A.~Lee, and G.~M. Whitesides,
\newblock Anal. Chem. {\bf 77}, 4726 (2005).

\bibitem{leslie2009}
D.~C. Leslie, C.~J. Easley, E.~Seker, J.~M. Karlinsey, M.~Utz, M.~R. Begley,
  and J.~P. Landers,
\newblock Nat. Phys. {\bf 5}, 231 (2009).

\bibitem{Hosoi2004}
A.~E. Hosoi and L.~Mahadevan,
\newblock Phys. Rev. Lett. {\bf 93}, 137802 (2004).

\bibitem{Weaver2010}
J.~A. Weaver, J.~Melin, D.~Stark, S.~R. Quake, and M.~A. Horowitz,
\newblock Nat. Phys. {\bf 6}, 218 (2010).

\bibitem{Mosadegh:2010gf}
B.~Mosadegh, C.-H. Kuo, Y.-C. Tung, Y.-s. Torisawa, T.~Bersano-Begey,
  H.~Tavana, and S.~Takayama,
\newblock Nat. Phys. {\bf 6}, 433 (2010).

\bibitem{forterre2005}
Y.~Forterre, J.~M. Skotheim, J.~Dumais, and L.~Mahadevan,
\newblock Nature {\bf 433}, 421 (2005).

\bibitem{Han2004}
J.~S. Han, J.~S. Ko, and J.~G. Korvink,
\newblock J. Micromech. Microengng {\bf 14}, 1585 (2004).

\bibitem{brinkmeyer2012}
A.~Brinkmeyer, M.~Santer, A.~Pirrera, and P.~M. Weaver,
\newblock Int. J. Solids Struct. {\bf 49}, 1077 (2012).

\bibitem{holmes2007}
D.~P. Holmes and A.~J. Crosby,
\newblock Adv. Mater. {\bf 19}, 3589 (2007).

\bibitem{Overvelde2015}
J.~T.~B. Overvelde, T.~Kloek, J.~J.~A. D'haen, and K.~Bertoldi,
\newblock Proc. Natl Acad. Sci. USA {\bf 112}, 10863 (2015).

\bibitem{Pandey2014}
A.~Pandey, D.~E. Moulton, D.~Vella, and D.~P. Holmes,
\newblock EPL {\bf 105}, 24001 (2014).

\bibitem{Gomez2017}
M.~Gomez, D.~E. Moulton, and D.~Vella,
\newblock Nat. Phys. {\bf 13}, 142 (2017).

\bibitem{Krylov2008}
S.~Krylov, B.~R. Ilic, D.~Schreiber, S.~Seretensky, and H.~Craighead,
\newblock J. Micromech. Microengng {\bf 18}, 055026 (2008).

\bibitem{Fargette2014}
A.~Fargette, S.~Neukirch, and A.~Antkowiak,
\newblock Phys. Rev. Lett. {\bf 112}, 137802 (2014).

\bibitem{Timoshenko}
S.~P. Timoshenko and J.~N. Goodier,
\newblock {\em Theory of Elasticity},
\newblock McGraw Hill, 1970.

\bibitem{Happel}
J.~Happel and H.~Brenner,
\newblock {\em Low {R}eynolds number hydrodynamics},
\newblock Kluwer, 1983.

\bibitem{Howell2009}
P.~Howell, G.~Kozyreff, and J.~Ockendon,
\newblock {\em Applied Solid Mechanics},
\newblock Cambridge University Press, 2009.

\bibitem{Kodio2017}
O.~Kodio, I.~M. Griffiths, and D.~Vella,
\newblock Phys. Rev. Fluids {\bf 2}, 014202 (2017).

\bibitem{Leal2007}
L.~G. Leal,
\newblock {\em Advanced Transport Phenomena: Fluid Mechanics and Convective
  Transport Processes},
\newblock Cambridge University Press, 2007.

\bibitem{Arena2017}
G.~Arena, R.~M.~J. Groh, A.~Brinkmeyer, R.~Theunissen, P.~M. Weaver, and  A.~Pirrera,
\newblock Proc. R. Soc. A  {\bf 473}, 20170334 (2017).

\bibitem{Cho2007}
H.~Cho, H.-Y. Kim, J.~Y. Kang, and T.~S. Kim,
\newblock J. Colloid Interf. Sci. {\bf 306}, 379 (2007).

\bibitem{Kim2009}
D.~Kim, Y.~W. Hwang, and S.-J. Park,
\newblock Microsyst. Technol. {\bf 15}, 919 (2009).


\bibitem{Thill:2008uk}
C.~Thill, J.~Etches, I.~Bond, K.~Potter, and P.~Weaver,
\newblock Aeronaut. J. {\bf 112}, 117 (2008).

\bibitem{endnote35}
See Supplementary Information (appended) for further details of experiments and theoretical calculations, which includes Refs \cite{Audoly2010,Ockendon1995,Patricio1998,Seydel2009}.

\bibitem{Audoly2010}
B.~Audoly and Y.~Pomeau,
\newblock {\em Elasticity and geometry: From hair curls to the non-linear response of shells.},
\newblock Oxford University Press, 2010.


\bibitem{Ockendon1995}
H.~Ockendon and J.~R.~Ockendon,
\newblock {\em Viscous Flow},
\newblock Cambridge University Press, 1995.

\bibitem{Patricio1998}
P.~Patricio, M.~Adda-Bedia, and M.~Ben Amar,
\newblock Physica D {\bf 124}, 285 (1998).


\bibitem{Seydel2009}
R.~Seydel,
\newblock {\em Practical bifurcation and stability analysis},
\newblock Springer, 2009.


\end{thebibliography}

\begin{thebibliography}{10}

\bibitem{audoly}
B.~Audoly and Y.~Pomeau.
\newblock {\em Elasticity and geometry: from hair curls to the non-linear
  response of shells}.
\newblock Oxford University Press, Oxford, 2010.

\bibitem{Gomez2017SI}
M.~Gomez, D.~E. Moulton, and D.~Vella.
\newblock Critical slowing down in purely elastic `snap-through' instabilities.
\newblock {\em Nat. Phys.}, 13:142--145, 2017.

\bibitem{howell}
P.~Howell, G.~Kozyreff, and J.~Ockendon.
\newblock {\em Applied {Solid} {Mechanics}}.
\newblock Cambridge University Press, Cambridge, 2009.

\bibitem{Kodio2017SI}
O.~Kodio, I.~M. Griffiths, and D.~Vella.
\newblock Lubricated wrinkles: imposed constraints affect the dynamics of
  wrinkle coarsening.
\newblock {\em Phys. Rev. Fluids}, 2:014202, 2017.

\bibitem{ockendon}
H.~Ockendon and J.~R. Ockendon.
\newblock {\em Viscous Flow}.
\newblock Cambridge University Press, Cambridge, 1995.

\bibitem{oh2006SI}
K.~W. Oh and C.~H. Ahn.
\newblock A review of microvalves.
\newblock {\em J. Micromech. Microengng}, 16(5):R13--R39, 2006.

\bibitem{Pandey2014SI}
A.~Pandey, D.~E. Moulton, D.~Vella, and D.~P. Holmes.
\newblock Dynamics of snapping beams and jumping poppers.
\newblock {\em EPL}, 105(2):24001, January 2014.

\bibitem{patricio1998}
P.~Patricio, M.~Adda-Bedia, and M.~B. Amar.
\newblock An elastica problem: instabilities of an elastic arch.
\newblock {\em Physica D: Nonlinear Phenomena}, 124(1-3):285--295, 1998.

\bibitem{seydel2009}
R. Seydel.
\newblock {\em Practical bifurcation and stability analysis}, volume~5.
\newblock Springer Science \& Business Media, 2009.

\bibitem{stone2004SI}
H.~A. Stone, A.~D. Stroock, and A.~Ajdari.
\newblock Engineering flows in small devices: microfluidics toward a
  lab-on-a-chip.
\newblock {\em Annu. Rev. Fluid Mech.}, 36:381--411, 2004.

\end{thebibliography}
\end{document}